\documentclass[prd,aps,preprintnumbers,floatfix,showpacs,byrevtex,twocolumn]{revtex4}
\usepackage{epsfig}
\usepackage{graphics}
\usepackage{amsmath}
\newcommand{\be}{\begin{equation}}
\newcommand{\ee}{\end{equation}}
\begin{document}

\preprint{
\vbox{
\hbox{ADP-01-63/T503}
\hbox{JLAB-THY-01-36}
}}

\title{Excited Baryons in Lattice QCD}

\author{W.~Melnitchouk$^{1,2}$, S.~Bilson-Thompson$^1$, F.D.R.~Bonnet$^1$,
        J.N.~Hedditch$^1$, F.X.~Lee$^{2,3}$, D.B.~Leinweber$^1$, 
        A.G.~Williams$^1$, J.M.~Zanotti$^1$ and J.B.~Zhang$^1$}

\affiliation{$^1$ Department of Physics and Mathematical Physics and\\
        Special Research Centre for the
        Subatomic Structure of Matter,                          \\
        University of Adelaide, 5005, Australia}

\affiliation{$^2$ Jefferson Lab, 12000 Jefferson Avenue,
        Newport News, VA 23606}

\affiliation{$^3$ Center for Nuclear Studies, Department of Physics,\\
        The George Washington University, Washington, D.C. 20052}

\begin{abstract}
  We present first results for the masses of positive and negative
  parity excited baryons calculated in lattice QCD using an ${\cal
    O}(a^2)$-improved gluon action and a fat-link irrelevant clover
  (FLIC) fermion action in which only the irrelevant operators are
  constructed with APE-smeared links.  The results are in agreement
  with earlier calculations of $N^*$ resonances using improved actions
  and exhibit a clear mass splitting between the nucleon and its
  chiral partner.  An correlation matrix analysis reveals
  two low-lying $J^P={1\over 2}^-$ states with a small mass splitting.
  The study of different $\Lambda$ interpolating fields suggests a
  similar splitting between the lowest two $\Lambda\frac{1}{2}^-$ octet
  states.  However, the empirical mass suppression of the
  $\Lambda^*(1405)$ is not evident in these quenched QCD simulations,
  suggesting a potentially important role for the meson cloud of the
  $\Lambda^*(1405)$ and/or a need for more exotic interpolating
  fields.
\end{abstract}

\vspace{3mm}

\maketitle

\newpage

\section{Introduction}

Understanding the dynamics responsible for baryon excitations provides
valuable insight into the forces which confine quarks inside baryons and
into the nature of QCD in the nonperturbative regime.
This is a driving force behind the experimental effort of the CLAS
Collaboration at Jefferson Lab, which is currently accumulating data of
unprecedented quality and quantity on various $N \to N^*$ transitions.
With the increased precision of the data comes a growing need to
understand the observed $N^*$ spectrum within QCD.
Although phenomenological low-energy models of QCD have been successful in
describing many features of the $N^*$ spectrum (for a recent review see
Ref.~\cite{CR}), they leave many questions unanswered, and calculations of
$N^*$ properties from first principles are indispensable.

One of the long-standing puzzles in spectroscopy has been the low mass
of the first positive parity excitation of the nucleon (the
$J^P={1\over 2}^+$ $N^*(1440)$ Roper resonance) compared with the lowest
lying odd parity excitation.
In a valence quark model, in a harmonic oscillator basis, the
${1\over 2}^-$ state naturally occurs below the $N=2$, ${1\over 2}^+$
state \cite{IK}.
Without fine tuning of parameters, valence quark models tend to leave
the mass of the Roper resonance too high.
Similar difficulties in the level orderings appear for the
${3\over 2}^+$ $\Delta^*(1600)$ and ${1\over 2}^+$ $\Sigma^*(1690)$,
which has led to speculations that the Roper resonances may be more
appropriately viewed as ``breathing modes'' of the states
\cite{BREATHE}, or described in terms of meson-baryon dynamics alone
\cite{FZJ}, or as hybrid baryon states with explicitly excited glue field
configurations \cite{LBL}.

Another challenge for spectroscopy is presented by the
$\Lambda^{1/2-}(1405)$, whose anomalously small mass has been
interpreted as an indication of strong coupled channel effects involving
$\Sigma\pi$, $\bar K N$, $\ldots$ \cite{L1405}, and a weak overlap with a
three-valence constituent-quark state.
In fact, the role played by Goldstone bosons in baryon spectroscopy has
received considerable attention recently \cite{MASSEXTR,YOUNG}.

It has been argued \cite{GBE} that a spin-flavour interaction associated
with the exchange of a pseudoscalar nonet of Goldstone bosons between
quarks can better explain the level orderings and hyperfine mass
splittings than the traditional (colour-magnetic) one gluon exchange
mechanism.
On the other hand, some elements of this approach, such as the
generalisation to the meson sector or consistency with the chiral
properties of QCD, remain controversial \cite{CR,ISGUR,TK}.
Furthermore, neither spin-flavour nor colour-magnetic interactions are able
to account for the mass splitting between the $\Lambda^{1/2^-}(1405)$ and
the $\Lambda^{3/2^-}(1520)$ (a splitting between these can arise in
constituent quark models with a spin-orbit interaction, however, this is
known to lead to spurious mass splittings elsewhere \cite{CR,CI}).
Recent work \cite{LARGE_NC} on negative parity baryon spectroscopy in the
large-$N_c$ limit has identified important operators associated with
spin-spin, spin-flavour and other interactions which go beyond the simple
constituent quark model, as anticipated by early QCD sum-rule analyses 
\cite{Leinweber:hh}.

The large number of states predicted by the constituent quark model and
its generalisations which have not been observed (the so-called
``missing'' resonances) presents another problem for spectroscopy.
If these states do not exist, this may suggest that perhaps a
quark--diquark picture (with fewer degrees of freedom) could afford a
more efficient description, although lattice simulation results provide
no evidence for diquark clustering \cite{DIQ}.
On the other hand, the missing states could simply have weak couplings to
the $\pi N$ system \cite{CR}.
Such a situation would present lattice QCD with a unique opportunity to
complement experimental searches for $N^*$'s, by identifying excited
states not easily accessible to experiment (as in the case of glueballs
or hybrids).

In attempting to answer these questions, one fact that will be clear
is that it is not sufficient to look only at the standard low mass
hadrons ($\pi, \rho, N$ and $\Delta$) on the lattice --- one must
consider the entire $N^*$ (and in fact the entire excited baryon)
spectrum.  In this paper we present the first results of octet baryon
mass simulations using an ${\cal O}(a^2)$ improved gluon action and an
improved Fat Link Irrelevant Clover (FLIC) \cite{FATJAMES} quark
action in which only the irrelevant operators are constructed using
fat links \cite{FATLINK}.  Configurations are generated on the Orion
supercomputer at the University of Adelaide. After reviewing in
Section II the main elements of lattice calculations of excited hadron
masses and a brief overview of earlier calculations, we describe in
Section III various features of interpolating fields used in this
analysis.  Section IV reviews the details of the lattice simulations,
and Section V gives an overview of the methodology for isolating
baryon resonance properties.  In Section VI we present the results
from our simulations and in Section VII make concluding remarks and
discuss possible future extensions of this work.

\section{Excited Baryons on the Lattice}

The history of excited baryons on the lattice is quite brief, although
recently there has been growing interest in finding new techniques to
isolate excited baryons, motivated partly by the experimental $N^*$
program at Jefferson Lab.
The first detailed analysis of the positive parity excitation of the
nucleon was performed by Leinweber \cite{LEIN1} using Wilson fermions and
an operator product expansion spectral ansatz.
DeGrand and Hecht \cite{DEGRAND92} used a wave function ansatz 
to access $P$-wave baryons, with Wilson fermions and relatively heavy
quarks.
Subsequently, Lee and Leinweber \cite{LL} introduced a parity projection
technique to study the negative parity ${1 \over 2}^-$ states using an
${\cal O}(a^2)$ tree-level tadpole-improved D$_{\chi 34}$ quark action,
and an ${\cal O}(a^2)$ tree-level tadpole-improved gauge action.
Following this, Lee \cite{LEE} reported results using a D$_{234}$ quark 
action with an improved gauge action on an anisotropic lattice to study 
the ${1 \over 2}^+$ and ${1 \over 2}^-$ excitations of the nucleon.
The RIKEN-BNL group \cite{DWF} has also performed an analysis of the 
$N^*({1\over 2}^-)$ and $N'({1\over 2}^+)$ excited states using domain
wall fermions.
More recently, a nonperturbatively improved clover quark action has
been used by Richards {\it et al.} \cite{RICHARDS} to study the
$N^*({1\over 2}^-)$ and $\Delta^*({3\over 2}^-)$ states, while
Nakajima {\it et al.} have studied the $N^*({1\over 2}^-)$ and
$\Lambda^*({1\over 2}^-)$ states using an anisotropic lattice with
an ${\cal O}(a)$ improved quark action \cite{Nakajima:2001js}.
Constrained-fitting methods based on Bayesian priors have also recently
been used by Lee {\em et al.} \cite{BAYESIAN} to study the two lowest
octet and decuplet positive and negative parity baryons using overlap
fermions with pion masses down to $\sim 180$~MeV.
While these authors claim to have observed the Roper in quenched QCD,
it remains to be demonstrated that this conclusion is
independent of the Bayesian-prior assumed in their
analysis\cite{LEIN1,ALLTON}.

Following standard notation, we define a two-point correlation function
for a spin-$\frac{1}{2}$ baryon $B$ as
\be
G_B(t,{\vec p})
\equiv \sum_{\vec{x}}\, e^{-i {\vec p} \cdot {\vec x}}
\left\langle \Omega \left| 
\chi_B(x)\bar\chi_B(0)
\right| \Omega \right\rangle
\label{2ptfunc}
\ee
where $\chi_B$ is a baryon interpolating field 
and where we have suppressed Dirac indices.
All formalism for correlation functions and
interpolating fields presented in this paper is carried
out using the Dirac representation of the $\gamma$-matrices.
The choice of interpolating field $\chi_B$ is discussed in
Section~III below.
The overlap of the interpolating field
$\chi_B$ with positive or negative parity states $| B^\pm \rangle$ is
parameterised by a coupling strength $\lambda_{B^\pm}$ which is
complex in general and which is defined by
\begin{subequations}
\begin{eqnarray}
\left\langle \Omega \left|\, \chi_B(0)\, \right| B^+ ,p,s \right\rangle
\!\!&=&\!\! \lambda_{B^+} \sqrt{M_{B^+} \over E_{B^+}}\ u_{B^+}(p,s)\, ,         \\
\left\langle \Omega \left|\, \chi_B(0)\, \right| B^- ,p,s \right\rangle
\!\!&=&\!\! \lambda_{B^-} \sqrt{M_{B^-} \over E_{B^-}}\, \gamma_5
u_{B^-}(p,s)\, ,
\end{eqnarray}
\end{subequations}
%
where $M_{B^\pm}$ is the mass of the state $B^\pm$,
$E_{B^\pm} = \sqrt{M_{B^\pm}^2 + {\vec p\,}^2}$ is its energy,
and $u_{B^\pm}(p,s)$ is a Dirac spinor with normalisation
$\overline{u}_{B^\pm}^{\alpha}(p,s)u_{B^\pm}^{\beta}(p,s) =
\delta^{\alpha\beta}$.
For large Euclidean time, the correlation function can be written as a
sum of the lowest energy positive and negative parity contributions
%
\begin{eqnarray}
\label{G+-}
G_B(t,{\vec p})
&\approx& \lambda_{B^+}^2
        { \left( \gamma \cdot p + M_{B^+} \right) \over 2 E_{B^+} }
        e^{- E_{B^+} \, t} \nonumber \\
 &+& \lambda_{B^-}^2
        { \left( \gamma \cdot p - M_{B^-} \right) \over 2 E_{B^-} }
        e^{- E_{B^-} \, t}\ ,
\label{cfunc}
\end{eqnarray}
%
when a fixed boundary condition in the time direction is used to 
remove backward propagating states.
The positive and negative parity states are isolated by taking the trace
of $G_B$ with the operator $\Gamma_+$ and $\Gamma_-$ respectively, where
\begin{eqnarray}
\Gamma_\pm &=&
{1\over 2} \left( 1 \pm {M_{B^\mp} \over E_{B^\mp}} \gamma_4 \right)\ .
\label{pProjOp}
\end{eqnarray}
For $\vec{p}=0$, $\Gamma_{\pm}^2 = \Gamma_{\pm}$ so that
$\Gamma_{\pm}$ are then parity projectors.
For $\vec p = 0$, the energy $E_{B^\pm} = M_{B^\pm}$ and using the
operator $\Gamma_\pm$ we can isolate the mass of the baryon $B^\pm$.
In this case, positive parity states propagate in the (1, 1) and (2, 2)
elements of the Dirac matrix of Eq.~(\ref{cfunc}), while negative parity 
states propagate in the (3, 3) and (4, 4) elements.

In terms of the correlation function $G_B$, the baryon effective mass
function is defined by
\be
M_B(t) = \log [G_B(t,\vec 0)] - \log [G_B(t+1,\vec 0)]\ .
\ee
Meson masses are determined via analogous standard procedures.

\section{Interpolating Fields}

In this analysis we consider two types of interpolating fields which have
been used in the literature.
The notation adopted is similar to that of Ref.~\cite{LWD}.
To access the positive parity proton we use as interpolating fields
\begin{eqnarray}
\label{chi1p}
\chi_1^{p +}(x)
&=& \epsilon_{abc}
\left( u^T_a(x)\ C \gamma_5\ d_b(x) \right) u_c(x)\ ,
\end{eqnarray}
and
\begin{eqnarray}
\label{chi2p}
\chi_2^{p +}(x)
&=& \epsilon_{abc}
\left( u^T_a(x)\ C\ d_b(x) \right) \gamma_5\ u_c(x)\ ,
\end{eqnarray}
where the fields $u$, $d$ are evaluated at Euclidean space-time point
$x$, $C$ is the charge conjugation matrix, $a, b$ and $c$ are
colour labels, and the superscript $T$ denotes the transpose.
These interpolating fields transform as a spinor under a parity
transformation.
That is, if the quark fields $q_a(x)\ (q=u,d, \cdots)$ transform as
$$
{\cal P} q_a(x) {\cal P}^\dagger = +\gamma_0 q_a(\tilde{x})\ ,
$$
where $\tilde{x} = (x_0 , -\vec{x})$, then
$$
{\cal P} \chi^{p +}(x) {\cal P}^\dagger
 = +\gamma_0 \chi^{p +}(\tilde{x})\ .
$$

For convenience, we introduce the shorthand notation
\begin{widetext}
\be
{\cal G}(S_{f_1},S_{f_2},S_{f_3}) \equiv
   \epsilon^{abc} \epsilon^{a'b'c'}
  \biggl\{ 
   S_{f_1}^{a a'}(x,0) \, {\rm tr} \left [ S_{f_2}^{b b' \, T}(x,0)
   S_{f_3}^{c c'}(x,0)
   \right ] + S_{f_1}^{a a'}(x,0) \, S_{f_2}^{b b'
     \, T}(x,0) \, S_{f_3}^{c c'}(x,0) \biggr \}\ ,
\label{F}
\ee
where $S^{a a'}_{f_{1-3}}(x,0)$ are the quark propagators in the
background link-field configuration $U$ corresponding to flavours
$f_{1-3}$.
This allows us to express the correlation functions in a compact form.
The associated correlation function for $\chi_1^{p+}$ can be
written as
\begin{equation}
G^{p+}_{11}(t,\vec p; \Gamma) = \left\langle \sum_{\vec x}
  e^{-i \vec p \cdot \vec x} {\rm tr}
  \left[ -\Gamma \, \, {\cal G}
    \left( S_u, \, \widetilde C S_d {\widetilde C}^{-1}, \, S_u \right)
  \right] \right\rangle\ ,
\label{p11CF}
\end{equation}
%
where $\langle\cdots\rangle$ is the ensemble average over the link fields,
$\Gamma$ is the $\Gamma_{\pm}$ projection operator from
Eq.~(\ref{pProjOp}), and $\widetilde C = C\gamma_5$.
For ease of notation, we will drop the angled brackets,
$\langle\cdots\rangle$, and all the following correlation functions will
be understood to be ensemble averages.  
For the $\chi_2^{p+}$ interpolating field, one can similarly write
%
\begin{equation}
G^{p+}_{22}(t,\vec p; \Gamma) = \sum_{\vec x}
  e^{-i \vec p \cdot \vec x} {\rm tr}
  \left[ - \Gamma \, \, {\cal G}
    \left( \gamma_5 S_u \gamma_5, \, \widetilde C S_d {\widetilde C}^{-1},
           \, \gamma_5 S_u \gamma_5
    \right)
  \right]\ ,
\label{p22CF}
\end{equation}
while the interference terms from these two interpolating fields are
given by, e.g.,
\begin{eqnarray}
G^{p+}_{12}(t,\vec p; \Gamma) 
&=& \sum_{\vec x} e^{-i \vec p \cdot \vec x} \hfill  \hfill
   {\rm tr} \biggl [ - \Gamma \, \, \biggl \{  {\cal G} \left ( S_u \gamma_5,\,
   \widetilde C S_d {\widetilde C}^{-1}, \,  S_u \gamma_5\right )
   \biggr \} \biggr ] . 
\label{p12CF} \\
G^{p+}_{21}(t,\vec p; \Gamma)
&=& \sum_{\vec x} e^{-i \vec p \cdot \vec x} \hfill  \hfill
   {\rm tr} \biggl [ - \Gamma \, \, \biggl \{  
   {\cal G} \left ( \gamma_5 S_u,\,
   \widetilde C S_d {\widetilde C}^{-1}, \,  \gamma_5 S_u
     \right ) \biggr \} \biggr ] . 
\label{p21CF}
\end{eqnarray}
\end{widetext}

The neutron interpolating field is obtained via the exchange 
$u\leftrightarrow d$, and the
strangeness --2, $\Xi$ interpolating field by
replacing the doubly represented $u$ or $d$ quark fields in
Eqs.~(\ref{chi1p}) and (\ref{chi2p}) by $s$ quark fields. 
$\Sigma$ and $\Xi$ interpolators are discussed in detail below.

As pointed out in Ref.~\cite{LEIN1}, because of the Dirac structure
of the ``diquark'' in the parentheses in Eq.~(\ref{chi1p}), in the
Dirac representation the field
$\chi_1^{p+}$ involves both products of
{\em upper}~$\times$~{\it upper}~$\times$~{\it upper} and
{\it lower}~$\times$~{\it lower}~$\times$~{\it upper} 
components of spinors for positive parity baryons, so that in
the nonrelativistic limit $\chi_1^{p+} = {\cal O}(1)$.
Here upper and lower refer to the large and small spinor components in
the standard Dirac representation of the $\gamma$ matrices. 
Furthermore, since the ``diquark'' couples to total spin 0, one expects
an attractive force between the two quarks, and hence better overlap with a lower energy
state than with a state in which two quarks do not couple to spin 0.

The $\chi_2^{p+}$ interpolating field, on the other hand, is known to
have little overlap with the nucleon ground state \cite{LEIN1,BOWLER}.
Inspection of the structure of the Dirac matrices in Eq.~(\ref{chi2p})
reveals that it involves only products of {\it upper}~$\times$~{\it
  lower}~$\times$~{\it lower} components for positive parity baryons,
so that $\chi_2^{p+} = {\cal O}(p^2 /E^2)$ vanishes in the
nonrelativistic limit.  As a result of the mixing of upper and lower
components, the ``diquark'' term contains a factor $\vec \sigma \cdot
\vec p$, meaning that the quarks no longer couple to spin 0, but are
in a relative $L=1$ state.  One expects therefore that two-point
correlation functions constructed from the interpolating field
$\chi_2^{p+}$ are dominated by larger mass states than those arising
from $\chi_1^{p+}$ at early Euclidean times.

While the masses of negative parity baryons are obtained directly 
from the (positive parity) interpolating fields in Eqs.~(\ref{chi1p}) and
(\ref{chi2p}) by using the parity projectors $\Gamma_\pm$, 
it is instructive nevertheless to examine the general properties
of the negative parity interpolating fields.
Interpolating fields with strong overlap with the negative parity
proton can be constructed by multiplying the previous positive parity
interpolating fields by
$\gamma_5$, $\chi_{1,2}^{p-} \equiv \gamma_5\ \chi_{1,2}^{p+}$.
In contrast to the positive parity case, both the interpolating fields
$\chi_1^{p -}$ and $\chi_2^{p -}$ mix upper and lower components, and
consequently both $\chi_1^{p -}$ and $\chi_2^{p -}$ are ${\cal O}(p/E)$.

Physically, two nearby $J^P = {1\over2}^-$ states are observed in the
excited nucleon spectrum.
In simple quark models, the splitting of these two orthogonal states is
largely attributed to the extent to which scalar diquark configurations
compose the wave function.
It is reasonable to expect $\chi_1^{p-}$ to have better overlap with
scalar diquark dominated states, and thus provide a lower effective mass
in the moderately large Euclidean time regime explored in lattice simulations.
If the effective mass associated with the $\chi_2^{p-}$ correlator is 
larger, then this would be evidence of significant overlap of 
$\chi_2^{p-}$ with the higher lying $N^{{1\over2}-}$ states.
In this event, a correlation matrix analysis (see Section V) will be
used to isolate these two states.

Interpolating fields for the other members of the flavour SU(3) octet
are constructed along similar lines.  For the positive parity $\Sigma^0$
hyperon one uses \cite{LWD}

\begin{eqnarray}
\chi_1^{\Sigma}(x)
= {1\over\sqrt{2}} \epsilon_{abc}
&\Big\{& \left( u^T_a(x)\ C \gamma_5\ s_b(x) \right) d_c(x)\nonumber \\
     &+& \left( d^T_a(x)\ C \gamma_5\ s_b(x) \right) u_c(x)
\Big\}\ , 
\label{chi1S} \\
\chi_2^{\Sigma}(x)
= {1\over\sqrt{2}} \epsilon_{abc}
&\Big\{& \left( u^T_a(x)\, C \, s_b(x) \right)\,\gamma_5 \,
  d_c(x)\nonumber \\
     &+& \left( d^T_a(x)\, C \, s_b(x) \right)\,\gamma_5\, u_c(x)
\Big\}\ .
\label{chi2S}
\end{eqnarray}
Interpolating fields used for accessing other charge states of
$\Sigma$ are obtained
by $d\rightarrow u$ or $u\rightarrow d$, producing correlation
functions analogous to those in Eqs.~(\ref{p11CF}) through (\ref{p12CF}).
Note that $\chi_1^{\Sigma}$ transforms as a triplet under SU(2) isospin.
An SU(2) singlet interpolating field can be constructed by replacing
$``+" \longrightarrow ``-"$ in Eqs.~(\ref{chi1S}) and (\ref{chi2S}).
For the SU(3) octet $\Lambda$ interpolating field (denoted by
``$\Lambda^8$''), one has
\begin{widetext}
\begin{eqnarray}
\chi_1^{\Lambda^8}(x)
&=& {1\over\sqrt{6}} \epsilon_{abc}
\Big\{ 2
   \left( u^T_a(x)\ C \gamma_5\ d_b(x) \right) s_c(x)
+ \left( u^T_a(x)\ C \gamma_5\ s_b(x) \right) d_c(x)\
- \left( d^T_a(x)\ C \gamma_5\ s_b(x) \right) u_c(x)
\Big\}\, , 
\label{chi1l8} \\
\chi_2^{\Lambda^8}(x)
&=& {1\over\sqrt{6}} \epsilon_{abc}
\Big\{2
   \left( u^T_a(x)\, C \, d_b(x) \right)\,\gamma_5 \, s_c(x)
+ \left( u^T_a(x)\, C \, s_b(x) \right)\,\gamma_5 \, d_c(x)\
- \left( d^T_a(x)\, C \, s_b(x) \right)\,\gamma_5 \, u_c(x)
\Big\}\, ,
\end{eqnarray}
which leads to the correlation function
\begin{eqnarray}
G^{\Lambda^8}_{11}(t,\vec p; \Gamma)  =
  {1 \over 6} \sum_{\vec x} e^{-i \vec p \cdot \vec x}
   {\rm tr} \biggl [ -\Gamma
\!\!  &\biggl \{& \!\!
     2 \,  {\cal G} \left ( S_s, \,\widetilde C S_u
   \widetilde C^{-1}, \, S_d \right )
   + 2 \, {\cal G} \left ( S_s, \, \widetilde C S_d
    \widetilde C^{-1}, \, S_u \right ) \nonumber\\
\!\! &+& \!\! 2 \, {\cal G} \left ( S_d, \, \widetilde C S_u
    \widetilde C^{-1}, \, S_s \right )
   + 2 \, {\cal G} \left ( S_u, \, \widetilde C S_d
    \widetilde C^{-1}, \, S_s \right ) \nonumber\\
\!\! &-& \!\! \phantom{2} \, {\cal G} \left ( S_d, \, \widetilde C S_s
    \widetilde C^{-1}, \, S_u \right )
   - \phantom{2} \, {\cal G} \left ( S_u, \, \widetilde C S_s
    \widetilde C^{-1}, \, S_d \right )
   \biggr \} \biggr ]\ 
\end{eqnarray}
%
and similarly for the correlation functions $G^{\Lambda^8}_{22}$,
$G^{\Lambda^8}_{12}$ and $G^{\Lambda^8}_{21}$.

The interpolating field for the SU(3) flavour singlet (denoted by
``$\Lambda^1$'') is given by \cite{LWD}
\begin{eqnarray}
\chi_1^{\Lambda^1}(x)
&=&  -2\ \epsilon_{abc}
\Big\{ -
   \left( u^T_a(x)\ C \gamma_5\ d_b(x) \right) s_c(x) 
+ \left( u^T_a(x)\ C \gamma_5\ s_b(x) \right) d_c(x)
- \left( d^T_a(x)\ C \gamma_5\ s_b(x) \right) u_c(x)
\Big\}\ , 
\label{chi1l1} \\
\chi_2^{\Lambda^1}(x)
&=& -2\ \epsilon_{abc}
\Big\{ -
   \left( u^T_a(x)\, C \, d_b(x) \right)\,\gamma_5 \, s_c(x) 
+\left( u^T_a(x)\, C \, s_b(x) \right)\,\gamma_5 \, d_c(x)
- \left( d^T_a(x)\, C \, s_b(x) \right)\,\gamma_5 \, u_c(x)
\Big\}\ ,
\end{eqnarray}
where the last two terms are common to both $\chi_1^{\Lambda^8}$ and
$\chi_1^{\Lambda^1}$.
The correlation function resulting from this field involves quite a few
terms,
%
\begin{eqnarray}
G^{\Lambda^{1}}_{11}(t,\vec p; \Gamma) &=&
   \epsilon^{abc}\epsilon^{a'b'c'}
   \sum_{\vec x} e^{-i \vec p \cdot \vec x} 
   {\rm tr} \biggl [ -\Gamma \, \, \biggl \{ 
   \gamma_5 S_s^{a a'} \, \widetilde C S_d^{c c' \, T} \widetilde C^{-1}
        \,  S_u^{b b'} \gamma_5
+  \gamma_5 S_u^{a a'} \, \widetilde C S_d^{c c' \, T} \widetilde C^{-1}
        \,  S_s^{b b'} \gamma_5 \nonumber \\
&+&  \gamma_5 S_s^{a a'} \, \widetilde C S_u^{c c' \, T} \widetilde C^{-1}
        \,  S_d^{b b'} \gamma_5 
+  \gamma_5 S_d^{a a'} \, \widetilde C S_u^{c c' \, T} \widetilde C^{-1}
        \,  S_s^{b b'} \gamma_5
+  \gamma_5 S_u^{a a'} \, \widetilde C S_s^{c c' \, T} \widetilde C^{-1}
        \,  S_d^{b b'} \gamma_5 \nonumber \\
&+&  \gamma_5 S_d^{a a'} \, \widetilde C S_s^{c c' \, T} \widetilde C^{-1}
        \,  S_u^{b b'} \gamma_5 
-  \gamma_5 S_s^{a a'} \gamma_5 \  {\rm tr} \left [
   S_d^{b b'} \, \widetilde C S_u^{c c' \, T} \widetilde C^{-1} \right ] \nonumber \\ 
&-&  \gamma_5 S_u^{a a'} \gamma_5 \  {\rm tr} \left [
   S_s^{b b'} \, \widetilde C S_d^{c c' \, T} \widetilde C^{-1} \right ]
-  \gamma_5 S_d^{a a'} \gamma_5 \  {\rm tr} \left [
   S_u^{b b'} \, \widetilde C S_s^{c c' \, T} \widetilde C^{-1} \right ]
   \biggr \} \biggr ]\ .
\end{eqnarray}

In order to test the extent to which SU(3) flavour symmetry is valid in
the baryon spectrum, one can construct another $\Lambda$ interpolating
field composed of the terms common to $\Lambda^1$ and $\Lambda^8$,
which does not make any assumptions about the SU(3) flavour symmetry
properties of $\Lambda$.
We define
\begin{eqnarray}
\chi_1^{\Lambda^c}(x)
&=& {1\over\sqrt{2}} \epsilon_{abc}
\Big\{
   \left( u^T_a(x)\ C \gamma_5\ s_b(x) \right) d_c(x)
- \left( d^T_a(x)\ C \gamma_5\ s_b(x) \right) u_c(x)
\Big\}\ , 
\label{chi1lc}\\
\chi_2^{\Lambda^c}(x)
&=& {1\over\sqrt{2}} \epsilon_{abc}
\Big\{
   \left( u^T_a(x)\, C \, s_b(x) \right)\,\gamma_5 \, d_c(x)
- \left( d^T_a(x)\, C \, s_b(x) \right)\,\gamma_5 \, u_c(x)
\Big\}\ ,
\label{chi2lc}
\end{eqnarray}
to be our ``common'' interpolating fields which are the isosinglet analog
of $\chi_1^\Sigma$ and $\chi_2^\Sigma$ in Eqs.~(\ref{chi1S}) and (\ref{chi2S}).
Such interpolating fields may be useful in determining the nature of the
$\Lambda^*(1405)$ resonance, as they allow for mixing between singlet and
octet states induced by SU(3) flavour symmetry breaking.
To appreciate the structure of the ``common'' correlation function, one
can introduce the function
%
\be
 \overline {\cal G}(S_{f_1},S_{f_2},S_{f_3}) =
   \epsilon^{abc} \epsilon^{a'b'c'}
\biggl \{ 
   S_{f_1}^{a a'}(x,0) \, {\rm tr} \left [ S_{f_2}^{b b' \, T}(x,0)
   S_{f_3}^{c c'}(x,0) \right ]
- S_{f_1}^{a a'}(x,0) \, S_{f_2}^{b b' \, T}(x,0) \,
   S_{f_3}^{c c'}(x,0) \biggr \}\ ,
\ee
which is recognised as ${\cal G}$ in Eq.~(\ref{F}) with the relative
sign of the two terms changed.
With this notation, the correlation function corresponding to the
$\chi_1^{\Lambda^c}$ interpolating field is
\be
 G_{11}^{\Lambda^c}(t,\vec p; \Gamma) =
   {1 \over 2} \sum_{\vec x} e^{-i \vec p \cdot \vec x}
   {\rm tr} \left [ -\Gamma \, \, \left \{ \overline {\cal G} \left ( S_d,
   \, \widetilde C S_s \widetilde C^{-1} ,\, S_u \right ) 
  + \overline {\cal G} \left ( S_u, \,
   \widetilde C S_s \widetilde C^{-1} ,\,
   S_d \right ) \right \} \right ]\ ,
\ee
and similarly for the correlation functions involving the
$\chi_2^{\Lambda^c}$ interpolating field.
\end{widetext}

\section{Lattice Simulations}

Having outlined the method of calculating excited baryon masses and the
choice of interpolating fields, we next describe the gauge and fermion
actions used in the present analysis.
Additional details of the simulations can be found in
Ref.~\cite{FATJAMES}.
%

\subsection{Gauge Action}

For the gauge fields, the Luscher-Weisz mean-field improved plaquette plus rectangle
action \cite{Luscher:1984xn} is used. We define
\begin{eqnarray}
S_{\rm G} \!&=&\! \frac{5\beta}{3}
      \sum_{\rm{sq}}\frac{1}{3}{\cal R}e\, {\rm{tr}}(1-U_{\rm{sq}}(x))
      \nonumber \\
    \!&-&\! \frac{\beta}{12u_{0}^2}
      \sum_{\rm{rect}}\frac{1}{3}{\cal R}e\, {\rm{tr}}(1-U_{\rm{rect}}(x))\, ,
\label{gaugeaction}
\end{eqnarray}
where the operators $U_{\rm{sq}}(x)$ and $U_{\rm{rect}}(x)$ are
defined as
\begin{subequations}
\begin{eqnarray}
U_{\rm{sq}}(x)
&=& U_{\mu}(x) U_{\nu}(x+\hat{\mu})
    U^{\dagger}_{\mu}(x+\hat{\nu}) U^\dagger_{\nu}(x)\ ,        \\
U_{\rm{rect}}(x)
&=& U_{\mu}(x) U_{\nu}(x+\hat{\mu})
    U_{\nu}(x+\hat{\nu}+\hat{\mu})                      \nonumber\\
& &\times U^{\dagger}_{\mu}(x+2\hat{\nu})
         U^{\dagger}_{\nu}(x+\hat{\nu})
         U^\dagger_{\nu}(x)                             \nonumber\\
&+& U_{\mu}(x) U_{\mu}(x+\hat{\mu})
    U_{\nu}(x+2\hat{\mu})                               \nonumber\\
& &\times U^{\dagger}_{\mu}(x+\hat{\mu}+\hat{\nu})
         U^{\dagger}_{\mu}(x+\hat{\nu})U^\dagger_{\nu}(x)\ .
\label{actioneqn}
\end{eqnarray}
\end{subequations}%
The link product $U_{\rm{rect}}(x)$ denotes the rectangular $1\times2$
and $2\times1$ plaquettes, and for the tadpole improvement factor we
employ the plaquette measure
\begin{equation}
u_0
= \left\langle \frac{1}{3}{\cal R}e \, {\rm{tr}}\langle U_{\rm{sq}}\rangle
  \right\rangle^{1/4}\ .
\label{uzero}
\end{equation}
Gauge configurations are generated using the Cabibbo-Marinari
pseudoheat-bath algorithm with three diagonal SU(2) subgroups looped
over twice.
Simulations are performed using a parallel algorithm with appropriate
link partitioning \cite{BONNET}.

The calculations of octet excited-baryon masses are performed on a
$16^3\times 32$ lattice at $\beta=4.60$.
The scale is set via the string tension obtained from the static quark
potential \cite{Edwards:1997xf}
$$
V({\rm \bf r}) = V_0 +\sigma r -e \left[\frac{1}{{\rm \bf r}}\right]
 + l\left(\left[\frac{1}{{\rm \bf r}}\right] - \frac{1}{r}\right)\ ,
$$
where $V_0$, $\sigma$, $e$ and $l$ are fit parameters, and
$\left[\frac{1}{{\rm \bf r}} \right]$ denotes the tree-level lattice
Coulomb term
$$
\left[\frac{1}{{\bf r}}\right]
= 4\pi\int\, \frac{d^3{\rm \bf k}}{(2\pi)^3}
  \cos({\rm \bf k}\cdot{\rm \bf r}) D_{44}(0,{\rm \bf k})\ ,
$$
with $D_{44}(k)$ the time-time component of the gluon propagator.
Note that $D_{44}(k_4,{\rm \bf k})$ is gauge-independent in the Breit
frame, $k_4 = 0$, since $k_4^2 / k^2=0$.
In the continuum limit, 
$$
\left[\frac{1}{{\rm \bf r}}\right]
\rightarrow \frac{1}{r}\ .
$$
Taking the physical value of the string tension to be
$\sqrt{\sigma} = 440$~MeV we find a lattice spacing of
$a = 0.122(2)$~fm.

\subsection{Fat-Link Irrelevant Fermion Action}

For the quark fields, we implement the Fat-Link Irrelevant Clover (FLIC)
action introduced in Ref.~\cite{FATJAMES}. 
Fat links 
are created by averaging or smearing links on the lattice with their
nearest transverse neighbours in a gauge covariant manner (APE smearing). In the
FLIC action, this reduces the problem of exceptional configurations
encountered with Wilson-style actions, and minimises the effect of
renormalisation on the action improvement terms.  By smearing only the
irrelevant, higher dimensional terms in the action, and leaving the
relevant dimension-four operators untouched, we retain short distance
quark and gluon interactions.  Furthermore, the use of fat links
\cite{FATLINK} in the irrelevant operators removes the need to fine
tune the clover coefficient in removing all ${\cal O}(a)$ artifacts.
It is now clear that FLIC fermions provide a new form of
nonperturbative ${\cal O}(a)$ improvement \cite{QNPproc}.

The smearing procedure \cite{APE} replaces a link, $U_{\mu}(x)$, with a
sum of the link and $\alpha$ times its staples
\begin{eqnarray}
&&\!\!U_{\mu}(x) \rightarrow U_{\mu}'(x) =
(1-\alpha) U_{\mu}(x) \\
&&\!\!+ \frac{\alpha}{6}\sum_{\nu=1 \atop \nu\neq\mu}^{4}
  \Big[ U_{\nu}(x)
        U_{\mu}(x+\nu a)
        U_{\nu}^{\dag}(x+\mu a)                         \nonumber \\
&&\!\!+ U_{\nu}^{\dag}(x-\nu a)
        U_{\mu}(x-\nu a)
        U_{\nu}(x-\nu a +\mu a)
  \Big] \,,  \nonumber
\end{eqnarray}
followed by projection back to SU(3).
We select the unitary matrix $U_{\mu}^{\rm FL}$ which maximises
$$
{\cal R}e \, {\rm{tr}}(U_{\mu}^{\rm FL}\, U_{\mu}^{' \dagger})\,
$$
by iterating over the three diagonal SU(2) subgroups of SU(3).
This procedure of smearing followed immediately by projection is repeated
$n$ times.
The fat links used in this investigation are created with $\alpha =
0.7$ and $n=4$ as discussed in Ref.~\cite{FATJAMES}.
The mean-field improved FLIC action is given by \cite{FATJAMES}
\begin{equation}
S_{\rm SW}^{\rm FL}
= S_{\rm W}^{\rm FL} - \frac{iC_{\rm SW} \kappa r}{2(u_{0}^{\rm FL})^4}\
             \bar{\psi}(x)\sigma_{\mu\nu}F_{\mu\nu}\psi(x)\ ,
\end{equation}
where $F_{\mu\nu}$ is constructed using fat links, and $u_{0}^{\rm
  FL}$ is calculated via Eq.~(\ref{uzero}) using the fat links. The
factor $C_{\rm SW}$ is the (Sheikholeslami-Wohlert) clover coefficient
\cite{CLOVER}, defined to be 1 at tree-level. The quark hopping
parameter is $\kappa = 1/(2m + 8r)$. We use the conventional choice
of the Wilson parameter, $r=1$.  The mean-field improved Fat-Link
Irrelevant Wilson action is
\begin{eqnarray}
S_{\rm W}^{\rm FL}
=  &\sum_x& \bar{\psi}(x)\psi(x) 
+ \kappa \sum_{x,\mu} \bar{\psi}(x)
    \bigg[ \gamma_{\mu}
      \bigg( \frac{U_{\mu}(x)}{u_0} \psi(x+\hat{\mu}) \nonumber \\
&-& \frac{U^{\dagger}_{\mu}(x-\hat{\mu})}{u_0} \psi(x-\hat{\mu})
      \bigg)
- r \bigg(
 \frac{U_{\mu}^{\rm FL}(x)}{u_0^{\rm  FL}} \psi(x+\hat{\mu})\nonumber\\
&+& \frac{U^{{\rm FL}\dagger}_{\mu}(x-\hat{\mu})}{u_0^{\rm FL}}
          \psi(x-\hat{\mu})
      \bigg)
    \bigg]\ .
\end{eqnarray}
Our notation for the fermion action uses the Pauli representation of
the Dirac $\gamma$-matrices defined in Appendix B of Sakurai
\cite{SAKURAI}.  In particular, the $\gamma$-matrices are Hermitian
with $\sigma_{\mu\nu} = [\gamma_{\mu},\ \gamma_{\nu}]/(2i)$.

As shown in Ref.~\cite{FATJAMES}, the mean-field
improvement parameter for the fat links is very close to 1, so that
the mean-field improved coefficient for $C_{\rm SW}$ is 
adequate \cite{FATJAMES}.
%
Another advantage is that one can now use highly improved definitions
of $F_{\mu\nu}$ (involving terms up to $u_0^{12}$), which give impressive
near-integer results for the topological charge \cite{SUNDANCE}.

%
In particular, we employ an ${\cal O}(a^4)$ improved definition of
$F_{\mu\nu}$ in which the standard clover-sum of four $1 \times 1$
loops lying in the $\mu ,\nu$ plane is combined with $2 \times 2$ and $3
\times 3$ loop clovers.
Bilson-Thompson {\it et al.} \cite{SUNDANCE} find
\begin{eqnarray}
F_{\mu\nu} = {-i\over{8}} \Big[\big( 
&&\!\! {3\over{2}}      W^{1 \times 1} -
 {3\over{20u_0^4}}W^{2 \times 2} \\
+&&\!\! {1\over{90u_0^8}}W^{3 \times 3}\big) - 
 {\rm h.c.}\Big]_{\rm traceless}\nonumber 
\end{eqnarray}
where $W^{n \times n}$ is the clover-sum of four $n \times n$ loops and
$F_{\mu\nu}$ is made traceless by subtracting $1/3$ of the trace from
each diagonal element of the $3 \times 3$ colour matrix.
This definition reproduces the continuum limit with ${\cal O}(a^6)$
errors.

A fixed boundary condition in the time direction is used for the fermions
by setting $U_t(\vec x, N_t) = 0\ \forall\ \vec x$ in the hopping terms
of the fermion action, with periodic boundary conditions imposed in the
spatial directions.
Gauge-invariant gaussian smearing \cite{Gusken:qx} in the spatial
dimensions is applied at the source to increase the overlap of the
interpolating operators with the ground states.
The source-smearing technique \cite{Gusken:qx} starts with a point
source, $\psi_0({\vec x}_0, t_0)$, at space-time location $({\vec
  x}_0, t_0) = (1,1,1,3)$ and proceeds via the iterative scheme,
\be
\psi_i(x,t) = \sum_{x'} F(x,x') \, \psi_{i-1}(x',t) \, ,
\ee
where
\begin{eqnarray}
F(x,x') &=& \frac{1}{(1+\alpha)} \Big( \delta_{x, x'} 
+ \frac{\alpha}{6} \sum_{\mu=1}^3 \big [ U_\mu(x) \, \delta_{x',
x+\widehat\mu} \nonumber \\
 &&\ \ + 
U_\mu^\dagger(x-\widehat\mu) \, \delta_{x', x-\widehat\mu} \big ]
\Big) \, .
\end{eqnarray}
Repeating the procedure $N$ times gives the resulting fermion field
\be
\psi_N(x,t) = \sum_{x'} F^N(x,x') \, \psi_0(x',t) \, .
\ee
The parameters $N$ and $\alpha$ govern the size and shape of the
smearing function and in our simulations we use $N=20$ and $\alpha=6$.

Five masses are used in the calculations \cite{FATJAMES} and the strange 
quark mass is taken to be the second heaviest quark mass in each case.
The analysis is based on a sample of 400 configurations, and the error
analysis is performed by a third-order, single-elimination jackknife,
with the $\chi^2$ per degree of freedom ($N_{\rm DF}$) obtained via
covariance matrix fits.

\section{Correlation Matrix Analysis}

In this section we outline the correlation matrix formalism for
calculations of masses, coupling strengths and optimal interpolating fields.
After demonstrating that the correlation functions are real, we proceed
to show how a matrix of such correlation functions may be used to isolate
states corresponding to different masses, and also to give information 
about the coupling of the operators to each of these states.

\subsection{The $U+U^{\ast}$ method}

A lattice QCD correlation function for the operator
$\chi_i \overline{\chi}_j$, where $\chi_i$ is the $i$-th interpolating
field for a particular baryon (e.g. $\chi_2^{p+}$ in Section~III), can be
written as
\begin{eqnarray}
\label{definition1}
{\cal G}_{ij} &\equiv& \left\langle\Omega|T(\chi_i \overline{\chi}_j)
  |\Omega\right\rangle \\
&=& \frac{\int {\cal D}U {\cal D}\bar\psi {\cal D}\psi
  e^{-S[U,\bar\psi,\psi]} \chi_i
\overline{\chi}_j}
{\int {\cal D}U {\cal D}\bar\psi{\cal D}\psi e^{-S[U,\bar\psi,\psi]}}\,,
\nonumber
\end{eqnarray} 
where spinor indices and spatial coordinates are suppressed for
ease of
notation.
The fermion and gauge actions can be separated such that
$ S[U,\bar\psi,\psi]=S_{G}[U] + \bar\psi M[U] \psi $.
Integration over the Grassmann variables $\bar\psi$ and $\psi$ then
gives
\begin{equation}
\label{naive}
{\cal G}_{ij}= \frac{ \int {\cal D}U e^{-S_{G}[U]} \det(M[U])H_{ij}[U] }
{\int {\cal D}U e^{-S_{G} [U]} \det(M[U])}\ ,
\end{equation}
where the term $H_{ij}$ stands for the sum of all full contractions of
$\chi_i \overline{\chi}_j$.
The pure gauge action $S_G$ and the fermion matrix $M$ satisfy
\begin{eqnarray}
S_G [U] = S_G [U^\ast]\ ,
\label{eqnarray:ActionProperties1}
\end{eqnarray}
and
\begin{eqnarray} 
T M[U^{\ast}] T^{-1} = M^{\ast}[U]\ ,
\label{eqnarray:ActionProperties2}
\end{eqnarray}
respectively, where $T$ is $C\gamma_5$ in the Sakurai convention,
adopted in Sec.~IV addressing the lattice actions.

Using the result of Eq.~(\ref{eqnarray:ActionProperties2}), one has
\begin{eqnarray}
\det\left(M[U^{\ast}]\right)
 &=& \det\left(M^{\ast}[U]\right)\ , 
\end{eqnarray}
and since ${\rm det}(M[U])$ is real,
\begin{equation}
\det\left(M[U^{\ast}]\right) = {\det\left(M[U]\right)}\ .
\end{equation}
Thus, $U$ and $U^{\ast}$ are configurations of equal weight in the measure
$\int {\cal D}U {\rm det}(M[U]){\rm exp}\left(-S_{G}[U]\right)$, in which
case ${\cal G}_{ij}$ can be written as
%
\begin{equation}
\label{endrel}
{\cal G}_{ij}
= \frac{1}{2}\left( \frac{\int {\cal D}U e^{-S_{G}[U]} \det(M[U])
\left\{H_{ij}[U] +H_{ij}[U^\ast]\right\} }
{\int {\cal D}U e^{-S_{G}[U]} \det(M[U]) }\right)\, .
\end{equation}
%

Let us define 
\be
G^{\pm}_{ij} \equiv {\rm tr_{sp}}\{\Gamma_{\pm}
{\cal G}_{ij}\} \, ,
\ee
where $\rm tr_{sp}$
denotes the spinor trace
and $\Gamma_{\pm}$ is the parity-projection operator defined
in Eq.~(\ref{pProjOp}).
If ${\rm tr_{sp}} \left\{ {\Gamma H_{ij}[U^\ast]} \right\}
  = {\rm tr_{sp}} \left\{ {\Gamma H_{ij}^{\ast}[U]} \right\}$,
then $G^{\pm}_{ij}$ is real. 
This can be shown by first noting that $H_{ij}$ will be products of
$\gamma$-matrices, fermion propagators, and link-field operators.
Fermion propagators have the form $M^{-1}$ and recalling that since
$T M[U^{\ast}] T^{-1}$=$M^{\ast}[U]$, then we have
$T M^{-1}[U^{\ast}] T^{-1}$=$(M^{-1}[U])^{\ast}$.
For products of link-field operators $O[U]$ contained in
$H_{ij}$, the condition
$O[U^{\ast}] = O^{\ast}[U]$ is equivalent to the requirement that the
coefficients of all link-products are real.
As long as this requirement is enforced, we can then simply proceed by
inserting $T T^{-1}$ inside the trace to show that the (spinor-traced)
correlation functions $G^{\pm}_{ij}$ are real.

In summary, provided that the products of link operators in the
interpolating fields are constructed using only real coefficients
then the correlation functions $G^{\pm}_{ij}$ are real.
This symmetry is explicitly implemented by 
including both $U$ and $U^*$ in the ensemble averaging
used to construct the lattice
correlation functions, providing an improved unbiased estimator which
is strictly real.
This is easily implemented at the correlation function level by
observing 
$$
M^{-1}(\{U_\mu^*\}) = [C\gamma_5 \, M^{-1}(\{U_\mu\})\,
(C\gamma_5)^{-1}]^*
$$ 
for quark propagators.

\subsection{Recovering masses, couplings and optimal interpolators}
\label{recover}

Let us again consider the momentum-space two-point function
for $t > 0$,
\begin{equation}
{\cal G}_{ij}(t,\vec{p}) = \sum_{\vec{x}} e^{-i\vec{p} \cdot \vec{x}}
\langle\Omega|
  \chi_i(t,\vec{x}) \overline{\chi}_j(0,\vec 0)
|\Omega\rangle\ .
\end{equation}
At the hadronic level,
\begin{eqnarray*}
{\cal G}_{ij}(t,\vec{p})\!\! &=&\!\! \sum_{\vec{x}} e^{-i\vec{p} \cdot \vec{x}}
\sum_{\vec{p}^{ \prime} ,s} \sum_{B}
\langle\Omega| \chi_i(t,\vec{x})| B ,p',s \rangle \nonumber \\
\!\!&\times&\!\!
\langle B ,p',s | \overline{\chi}_j(0,\vec 0) |\Omega\rangle\ ,
\end{eqnarray*}
where the $|B ,p',s\rangle$ are a complete set of states with momentum
$p'$ and spin $s$
%
\begin{equation}
\sum_{\vec{p}^{\prime}} \sum_{B} \sum_{s}
 |B ,p',s\rangle\langle B ,p',s|=I\ .
\end{equation}
We can make use of translational invariance to write
\begin{widetext}
\begin{eqnarray}
{\cal G}_{ij}(t,\vec{p})
&=& \sum_{\vec{x}} e^{-i\vec{p} \cdot \vec{x}}
\sum_{\vec{p}^{\prime}} \sum_{s} \sum_{B}
\left\langle\Omega\left|
  e^{\hat{H}t} e^{-i\hat{\vec{P}} \cdot \vec{x}}\chi_i(0)
  e^{i\hat{\vec{P}} \cdot \vec{x}} e^{-\hat{H}t}
\right| B ,p',s \right\rangle
\left\langle B ,p',s \left| \overline{\chi}_j(0)
\right| \Omega \right\rangle                            \nonumber \\
&=& \sum_{s} \sum_{B} e^{-{E_{B}t}}
\left\langle\Omega| \chi_i(0) |B ,p,s \rangle
 \langle B ,p,s | \overline{\chi}_j(0)|\Omega\right\rangle\ .
\end{eqnarray}
\end{widetext}

It is convenient in the following discussion to label the states
which have the $\chi$ interpolating field quantum numbers and
which survive the parity projection as $|B_{\alpha}\rangle$ for
$\alpha=1,2,\cdots,N$.  In general the number of states,
$N$, in this tower of excited states may be infinite, but we
will only ever need to consider a finite set of the lowest such
states here.
After selecting zero momentum, $\vec p=0$, the parity-projected trace
of this object is then
\begin{equation}
\label{eqn:Gijequation}
G^{\pm}_{ij}(t)
 = {\rm tr_{sp}}\{\Gamma_{\pm} {\cal G}_{ij}(t,\vec 0)\}
 = \sum_{\alpha=1}^{N} e^{-{m_{\alpha}}t}
   \lambda^{\alpha}_i \overline{\lambda}^{\alpha}_j\ ,
\end{equation}
where $\lambda^{\alpha}_i$ and $\overline{\lambda}^{\alpha}_j$ are
coefficients denoting the
couplings of the interpolating fields $\chi_i$ and $\overline{\chi}_j$,
respectively, to the state $\left|B_{\alpha}\right\rangle$.
If we use identical source and sink interpolating fields then it
follows from the definition of the coupling strength that
$\overline{\lambda}^{\alpha}_j = (\lambda^{\alpha}_j)^*$
and from Eq.~(\ref{eqn:Gijequation}) we see that
$G^{\pm}_{ij}(t)=[G^{\pm}_{ji}(t)]^*$, i.e., $G^\pm$ is
a Hermitian matrix.  If, in addition, we use only real
coefficients in the link products, then $G^\pm$ is a real
symmetric matrix.  For the correlation matrices that we
construct we have real link coefficients but we use smeared
sources and point sinks and so in our calculations $G$ is a real
but non-symmetric matrix.  Since $G^\pm$ is a real matrix for the
infinite number of possible choices of interpolating fields with
real coefficients, then we can take
$\lambda^{\alpha}_i$ and $\overline{\lambda}^{\alpha}_j$ to
be real coefficents here without loss of generality.

Suppose now that we have $M$ creation and annihilation operators,
where $M < N$.
We can then form an $M \times M$ approximation to the full $N \times N$
matrix $G$.
At this point there are two options for extracting masses.
The first is the standard method for calculation of effective masses
at large $t$ described in Section~II.
%
%
The second option is to extract the masses through a correlation-matrix
procedure~\cite{MCNEILE}.

Let us begin by considering the ideal case where we have $N$
interpolating fields with the same quantum numbers, but which
give rise to $N$ linearly independent states when acting on the
vacuum.  In this case we can constuct $N$ ideal interpolating
source and sink fields which perfectly isolate the $N$ individual
baryon states $|B_\alpha\rangle$, i.e.,
\begin{subequations}
\begin{eqnarray}
\overline{\phi}^{\alpha} &=& \sum_{i=1}^N 
          u^{\alpha}_i\ \overline{\chi}_i\ , \\
\phi^{\alpha} &=& \sum_{i=1}^N v^{\ast \alpha}_i\ \chi_i\ ,
\end{eqnarray}
\label{lincomIF}
\end{subequations}%
such that
\begin{subequations}
\begin{eqnarray}
\left\langle B_{\beta}\right| \overline{\phi}^{\alpha}
\left| \Omega\right\rangle
&=& \delta_{\alpha\beta}\ \overline{z}^{\alpha}\
\overline{u}(\alpha,p,s)\ ,                     \\
\left\langle \Omega \right | \phi^{\alpha}
\left| B_{\beta}\right\rangle
&=& \delta_{\alpha\beta}\ z^{\alpha}\
u(\alpha,p,s)\ ,
\end{eqnarray}
\label{PhiExpression}
\end{subequations}%
where $z^\alpha$ and $\overline{z}^\alpha$ are the coupling strengths
of $\phi^\alpha$ and $\overline{\phi}^\alpha$ to the state
$|B_\alpha\rangle$.
The coefficients $u_i^\alpha$ and $v_i^{\ast \alpha}$ in
Eqs.~(\ref{lincomIF}) may differ when the source and sink have different
smearing prescriptions, again indicated by the differentiation between
$z^\alpha$ and $\overline{z}^\alpha$.  
For notational convenience for the remainder of this
discussion repeated indices $i,j,k$ are to be understood
as being summed over. 
At $\vec{p}=0$, it follows that,
\begin{eqnarray}
\label{ActingLeft}
G^{\pm}_{ij}(t)\ u^{\alpha}_j
&=& \left(\sum_{\vec{x}} {\rm tr_{sp}}
\left\{ \Gamma_{\pm}
  \left \langle \Omega \right |
  \chi_i \overline{\chi}_j
  \left| \Omega \right\rangle
\right\}
\right) u^{\alpha}_j                    \nonumber\\
&=& \lambda^{\alpha}_i \overline{z}^{\alpha} 
    e^{-m_{\alpha} t} .
\end{eqnarray}
The only $t$-dependence in this expression comes from the exponential
term, which leads to the recurrence relationship
\begin{equation}
G^{\pm}_{ij} (t) \, u^{\alpha}_j
= e^{m_{\alpha}} G^{\pm}_{ik} (t+1) \, u^{\alpha}_k\ ,
\label{eveqn}
\end{equation}
which can be rewritten as
\begin{equation}
[G^{\pm} (t+1)]_{ki}^{-1} G^{\pm}_{ij} (t) \, u^{\alpha}_j
= e^{m_{\alpha}}\, u^{\alpha}_k\ .
\label{eveqn2}
\end{equation}
This is recognized as an eigenvalue equation for the matrix
$[G^{\pm} (t+1)]^{-1} G^{\pm}(t)$ with 
eigenvalues $e^{m_{\alpha}}$ and eigenvectors $u^\alpha$.
Hence the natural logarithms of the eigenvalues of
$[G^{\pm} (t+1)]^{-1} G^{\pm}(t)$ are the masses of the
$N$ baryons in the tower of excited states corresponding
to the selected parity and the quantum numbers of the
$\chi$ fields.  The eigenvectors are the coefficients of the
$\chi$ fields providing the ideal linear combination for that
state.  Note that since here we use only real coefficients in our
link products, then $[G^{\pm} (t+1)]^{-1} G^{\pm}(t)$ is a real
matrix and so $u^\alpha$ and $v^\alpha$ will be real eigenvectors.
It also then follows that $z^{\alpha}$ and $\overline{z}^\alpha$
will be real.
These coefficients are examined in detail in the following
section.

One can also construct the equivalent left-eigenvalue equation to recover
the $v$ vectors, providing the optimal linear combination of
annihilation interpolators,
\begin{equation}
v^{\ast \alpha}_k G^{\pm}_{kj} (t)
= e^{m_{\alpha}} v^{\ast \alpha}_i G^{\pm}_{ij} (t+1)\ .
\end{equation}
Recalling Eq.~(\ref{ActingLeft}), one finds:
\begin{eqnarray}
G^{\pm}_{ij} (t)\ u^{\alpha}_j
&=& \overline{z}^{\alpha} \lambda^{\alpha}_i
    e^{-m_{\alpha} t}\ ,                                   \\
v^{\ast \alpha}_i\ G^{\pm}_{ij} (t)
&=& {z}^{\alpha} \overline{\lambda}^{\alpha}_j e^{-m_{\alpha} t }\ , \\
v^{\ast \alpha}_k\ G^{\pm}_{kj} (t) G^{\pm}_{il} (t)\ u^{\alpha}_l
&=& z^{\alpha} \overline{z}^{\alpha} \lambda^{\alpha}_i 
\overline{\lambda}^{\alpha}_j e^{-2m_{\alpha} t}\ .
\label{preprojection}
\end{eqnarray}
The definitions of Eqs.~(\ref{PhiExpression}) imply
\begin{equation}
v^{\ast \alpha}_i\ G^{\pm}_{ij}(t)\ u^{\alpha}_j =
z^{\alpha}\overline{z}^{\alpha} e^{-m_{\alpha} t } ,
\label{projection}
\end{equation}
indicating the eigenvectors may be used to construct a correlation
function in which a single state mass $m_\alpha$is isolated
and which can be analysed
using the methods of Section~II. We refer to this as the projected
correlation function in the following.
Combining Eqs.~(\ref{preprojection}) and (\ref{projection}) leads us to
the result
\begin{equation}
\frac{v^{\ast \alpha}_k \ G_{kj}(t) G_{il}(t)\ u^{\alpha}_l}
{v^{\ast \alpha}_k G_{kl}(t) u^{\alpha}_l }
= \lambda^{\alpha}_{i}\overline{\lambda}^{\alpha}_{j}
  e^{-m_{\alpha} t} 
\ .
\label{eqn:ratios}
\end{equation}
By extracting all $N^2$ such ratios, we can exactly
recover all of the real couplings $\lambda^{\alpha}_{i}$ and 
$\overline{\lambda}^{\alpha}_{j}$ of 
$\chi_i$ and $\overline{\chi}_j$ respectively to
the state $|B_\alpha\rangle$.
Note that throughout this section no assumptions have been made about
the symmetry properties of $G_{ij}^{\pm}$. This is essential due to our
use of smeared sources and point sinks.

In practice we will only have a relatively small number, $M<N$,
of interpolating fields in any given analysis.  These $M$
interpolators should be chosen to have good overlap with the
lowest $M$ excited states in the tower and we should attempt
to study the ratios in Eq.~(\ref{eqn:ratios}) at early to
intermediate Euclidean times, where the contribution of the
$(N-M)$ higher mass states will be suppressed but where
there is still sufficient signal to allow the lowest $M$ states
to be seen.  This procedure will lead
to an estimate for the masses of each of the lowest $M$ states
in the tower of excited states.  Of these $M$
predicted masses, the highest will in general have the largest
systematic error while the lower masses will be
most reliably determined.  Repeating the analysis with varying
$M$ and different combinations of interpolating fields
will give an objective measure of the reliability of the
extraction of these masses.

In our case of a modest $2 \times 2$ correlation matrix ($M=2$)
we take a cautious approach to the selection of the
eigenvalue analysis time.
As already explained, we perform the eigenvalue analysis at an early
to moderate Euclidean time where statistical noise is suppressed
and yet contributions from at least the lowest two mass states
is still present.  One must exercise caution in performing the
analysis at too early a time, as more than the desired
$M=2$ states may be contributing to the $2\times 2$ matrix of
correlation functions.

We begin by projecting a particular parity, and then investigate the
effective mass plots of the elements of the correlation matrix.  Using
the covariance-matrix based $\chi^2/N_{\rm DF}$, we identify the time
slice at which all correlation functions of the correlation matrix are
dominated by a single state.  In practice, this time slice is
determined by the correlator providing the lowest-lying effective mass
plot. 
The eigenvalue analysis is performed at one time slice earlier,
thus ensuring the presence of multiple states in the elements of the
correlation function matrix, minimising statistical uncertainties,
and hopefully providing a clear signal for the analysis.  
In this approach minimal new
information has been added, providing the best opportunity that the $2
\times 2$ correlation matrix is indeed dominated by 2 states.  The
left and right eigenvectors are determined and used to project
correlation functions containing a single state from the correlation
matrix as indicated in Eq.~(\ref{projection}).  These correlation
functions are then subjected to the same covariance-matrix based
$\chi^2/N_{\rm DF}$ analysis to identify new acceptable fit windows for
determining the masses of the resonances.

\section{Results}

\subsection{Effective masses and the correlation matrix}
\label{6a}

The correlation matrix analysis has a significant impact on the
resolution of states obtained with the $\Lambda^c$ interpolating
fields of Eqs.~(\ref{chi1lc}) and (\ref{chi2lc}). Hence we begin our
discussion with a focus on these correlation functions.

\begin{figure}[t] 
\begin{center}
\rotatebox{90}{\epsfig{file=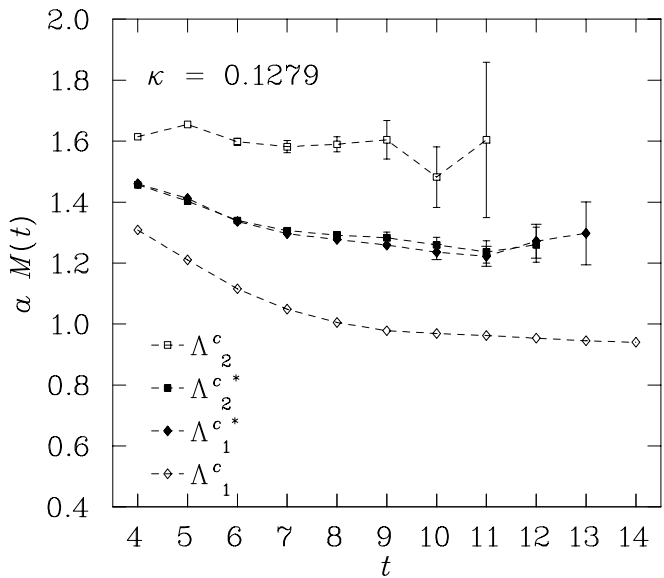,height=\hsize}} 
\vspace*{0.5cm}
\caption{Effective masses of the lowest lying positive and negative
  parity $\Lambda$ states obtained using the $\Lambda^c$ interpolating
  field from 400 configurations using the FLIC action defined with 4
  sweeps of smearing at $\alpha=0.7$.  The $J^P = {1\over 2}^+$
  (${1\over 2}^-$) states labeled $\Lambda^c_1$ ($\Lambda^{c*}_1$) and
  $\Lambda^c_2$ ($\Lambda^{c*}_2$) are obtained using the $\chi_1 \overline
  \chi_1$ and $\chi_2 \overline \chi_2$ interpolating fields,
  respectively.  The smeared source is at $t=3$.}
\end{center}
\end{figure}

\begin{figure}[t] 
\begin{center}
\rotatebox{90}{\epsfig{file=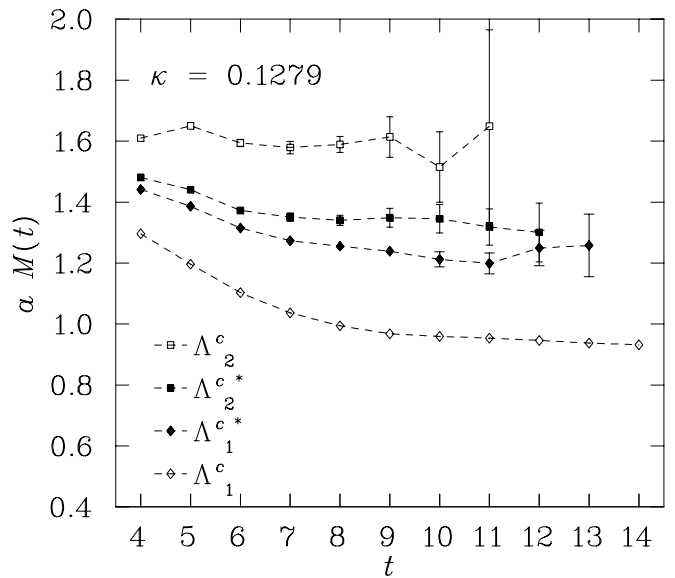,height=\hsize}} 
\vspace*{0.5cm}
\caption{As in Fig.~1, but for states obtained using the correlation
  functions projected from the correlation matrix as in
  Eq.~(\protect{\ref{projection}}).
  \label{new}}
\end{center}
\end{figure}

The effective mass plots for the positive and negative parity $\Lambda$
states obtained using the $\Lambda^c$ interpolating field
in the $\chi_1 \overline \chi_1$ and $\chi_2
\overline \chi_2$ correlation functions are shown in Fig.~1 for the
FLIC action.  Good values of the covariance matrix based $\chi^2 / N_{\rm
DF}$ are obtained for the ground state $(\Lambda^c_1)$ for many
different time-fitting intervals as long as one fits after time slice
9.  Similarly, the lowest $J^P = {1\over 2}^-$ excitation for the
$\chi_1 \overline \chi_1$ correlator $(\Lambda^{c*}_1)$ requires fits following
time slice 8.  
The ground state $(\Lambda^c_1)$ mass obtained
from $\chi_1 \overline \chi_1$ alone uses time slices 10--14 while the
first odd-parity excited state, $(\Lambda^{c*}_1)$ uses time slices 9--12.  The
states obtained from the $\chi_2 \overline \chi_2$ correlation
function plateau at earlier times and are also subject to
noise earlier in time than the states obtained with the $\chi_1
\overline \chi_1$ correlator.  For these reasons, good values of
$\chi^2 / N_{\rm DF}$ are obtained on the time interval 6--8 for the
positive parity states ($\Lambda^c_2$), and time interval 8--11 for the
negative parity states ($\Lambda^{c*}_2$).
Hence, the time slice at which the eigenvalue analysis of the
correlation matrix is performed is at $T=9$ for the even-parity
pair of states and at $T=8$ for the odd-parity pair of states.
Selecting only one time slice earlier than that allowed by $\chi^2$
considerations provides the best chance 
that only two states are present in the correlation matrix at that
time.

To guarantee the robustness of the eigenvector analysis and the
subsequent projection procedure, various consistency checks are made
at each stage of the process.  For instance, a check is made to
determine that the eigenvalue in Eq.~(\ref{eveqn}) is positive, and
that the mass determined from the projected correlation function
defined in Eq.~(\ref{projection}) is within the statistical
fluctuations of the mass extracted without this analysis.  
For the octet interpolating fields, off-diagonal elements are often
suppressed by an order of magnitude relative to the diagonal elements
and statistical noise can prevent the eigenvalue analysis from being
successful. However, the strong suppression of off-diagonal elements
is a clear signature that the mixing of the interpolating
fields in these states is negligible.

When the consistency checks are not satisfied, we have explored the
possibility of stepping back to the previous time slice and performing
the correlation matrix analysis there.  In some cases, the mass of the
lower-lying state reliably obtained via Euclidean time evolution is
seen to increase in the eigenvalue analysis, indicating a failure of
the correlation matrix analysis. The increase in the eigenvalue
indicates that there are significant contributions from three or more
states in the $2\times 2$ correlation matrix, thus spoiling the
possibility of successful state isolation. In this case, the
correlation matrix analysis is unable to provide additional
information and masses are
reported from the $\chi_1\overline\chi_1$ or $\chi_2\overline\chi_2$
correlators as appropriate.

\begin{figure}[t] 
\begin{center}
\rotatebox{90}{\epsfig{file=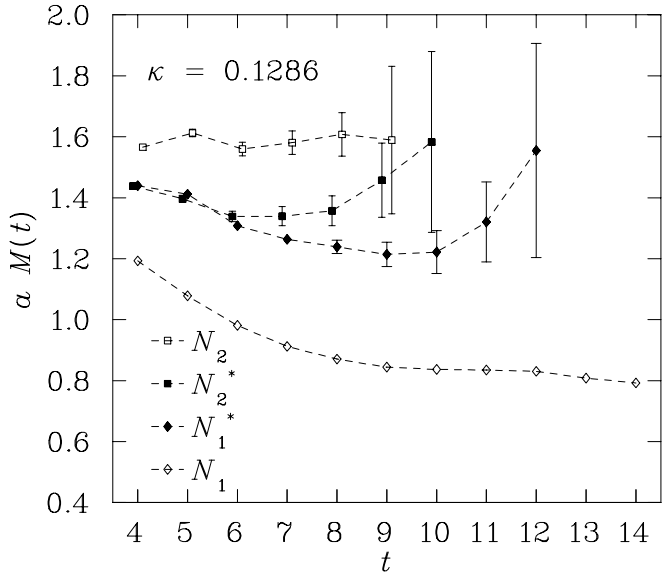,height=\hsize}} 
\vspace*{0.35cm}
\caption{As in Fig.~1, but for the nucleon states obtained using the correlation
  functions defined in  Eqs.~(\protect{\ref{p11CF}}) and
  (\protect{\ref{p22CF}}). 
  \label{Neff}}
\end{center}
\end{figure}

\begin{figure}[t] 
\begin{center}
\rotatebox{90}{\epsfig{file=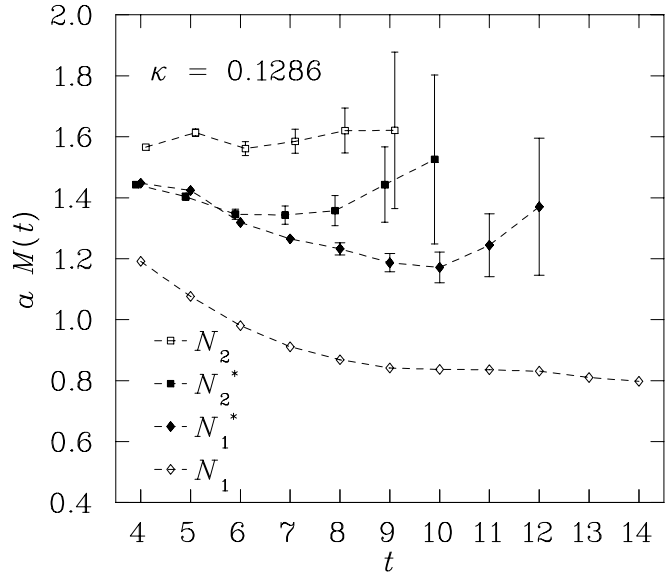,height=\hsize}} 
\vspace*{0.5cm}
\caption{As in Fig.~2, but for the nucleon states obtained using the correlation
  functions projected from the correlation matrix as in
  Eq.~(\protect{\ref{projection}}).
  \label{NeffCM}}
\end{center}
\end{figure}

\begin{figure*}[t!] 
\begin{center}
\leavevmode
\epsfig{figure=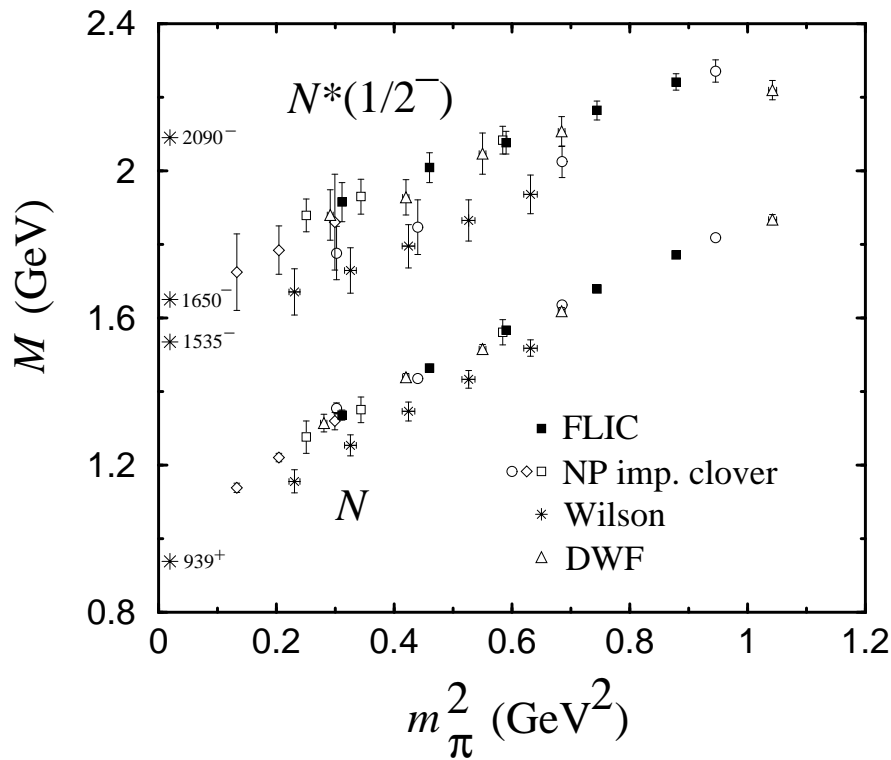,height=10cm}
\vspace*{0.5cm}
\caption{Masses of the nucleon ($N$) and the lowest $J^P={1\over 2}^-$
  excitation (``$N^*$'').  The FLIC and Wilson results are from the
  present analysis, with the DWF \protect\cite{DWF} and NP improved
  clover \protect\cite{RICHARDS} results shown for comparison.  The
  empirical nucleon and low lying $N^*({1\over 2}^-)$ masses are
  indicated by the asterisks along the ordinate.}
\end{center}
\end{figure*}
\begin{table*} 
\begin{center}
\caption{Values of $\kappa$ used in this analysis and the corresponding
        pion and nucleon resonance masses for the FLIC
        action with 4 sweeps of smearing at $\alpha=0.7$.
        Here $\kappa_{\rm cr} = 0.1300$, and a string tension analysis
        gives $a=0.122(2)$ fm for $\sqrt\sigma=440$ MeV.
\label{kappa}}
\begin{ruledtabular}
  \begin{tabular}{cccccc}
$\kappa$   & $m_{\pi} a$  & $m_{N_1} a$   & $m_{N^*_1} a$ & $m_{N^*_2} a$ & $m_{N_2} a$ \\ \hline
0.1260 & 0.5807(18) & 1.0972(49) & 1.388(14) & 1.442(12) & 1.676(12) \\ 
0.1266 & 0.5343(19) & 1.0400(53) & 1.340(16) & 1.404(15) & 1.642(13) \\
0.1273 & 0.4758(21) & 0.9701(59) & 1.286(19) & 1.363(20) & 1.605(15) \\
0.1279 & 0.4203(23) & 0.9067(67) & 1.244(25) & 1.345(29) & 1.580(18) \\
0.1286 & 0.3457(28) & 0.8273(86) & 1.186(33) & 1.374(57) & 1.571(26) \\
\end{tabular}
\end{ruledtabular}
\end{center}
\end{table*}

Figure~\ref{new} illustrates the effective mass plots of the
correlation functions projected from the correlation matrix as in
Eq.~(\ref{projection}).  The improved plateau behaviour is readily
visible.  Whereas in Fig.~1 the odd-parity effective masses are
crossing at $t=6$ and have minimal mass splitting, significant mass
splitting between the two states is already apparent at $t=6$ in
Fig.~\ref{new}.  The covariance based $\chi^2 / N_{\rm DF}$ indicates
that acceptable plateaus in the effective mass plots start even
earlier in some cases.  The increase in mass splitting between the two
negative parity states is more dramatic for $\Lambda^*_c$ than for the
octet baryon interpolating fields. There the off diagonal elements of
the correlation matrix are suppressed for the negative parity octet
baryons, but not so for $\Lambda^*_c$. As a result, the projection of
states has only only a small effect for the octet baryon interpolators
and this is detailed in Section~VIc.

Figs.~\ref{Neff} and \ref{NeffCM} show the effective mass plots of the
nucleon correlation functions $\chi_1 \overline \chi_1$ and $\chi_2
\overline \chi_2$ and following projection of the correlation matrix,
respectively. Plots for the lightest quark mass considered are
presented. The covariance matrix analysis of all quark masses
indicates the following analysis windows in Euclidean time,
\begin{eqnarray*}
 N_1 \, , 10-14 &;& N_1^* \, , 9-12 \, ; \\
 N_2^* \, , 8-11 &;& N_2 \, , 6-8 \, . 
\end{eqnarray*}

A comparison of Figs.~\ref{Neff} and \ref{NeffCM} indicates that the
correlation matrix analysis has a significantly smaller effect for the
nucleon interpolators than the $\Lambda_c$ interpolators.  This
suggests that the states created by the interpolating fields
$\chi_1$ and $\chi_2$ have
good overlap with the two lowest-lying physical nucleon states.

\subsection{Resonance masses and lattice action dependence}

In Fig.~5 we show the nucleon and $N^*({1\over 2}^-)$ masses as a
function of the pseudoscalar meson mass squared, $m_\pi^2$.  The
results of the new simulations are
indicated by the filled squares for the FLIC action, and by the stars
for the Wilson action (the Wilson points are obtained from a sample of
50 configurations).
The values of $m_\pi^2$ correspond to $\kappa$ values given in
Table~\ref{kappa}.

\begin{figure*}[t!] 
\begin{center}
\leavevmode
\epsfig{figure=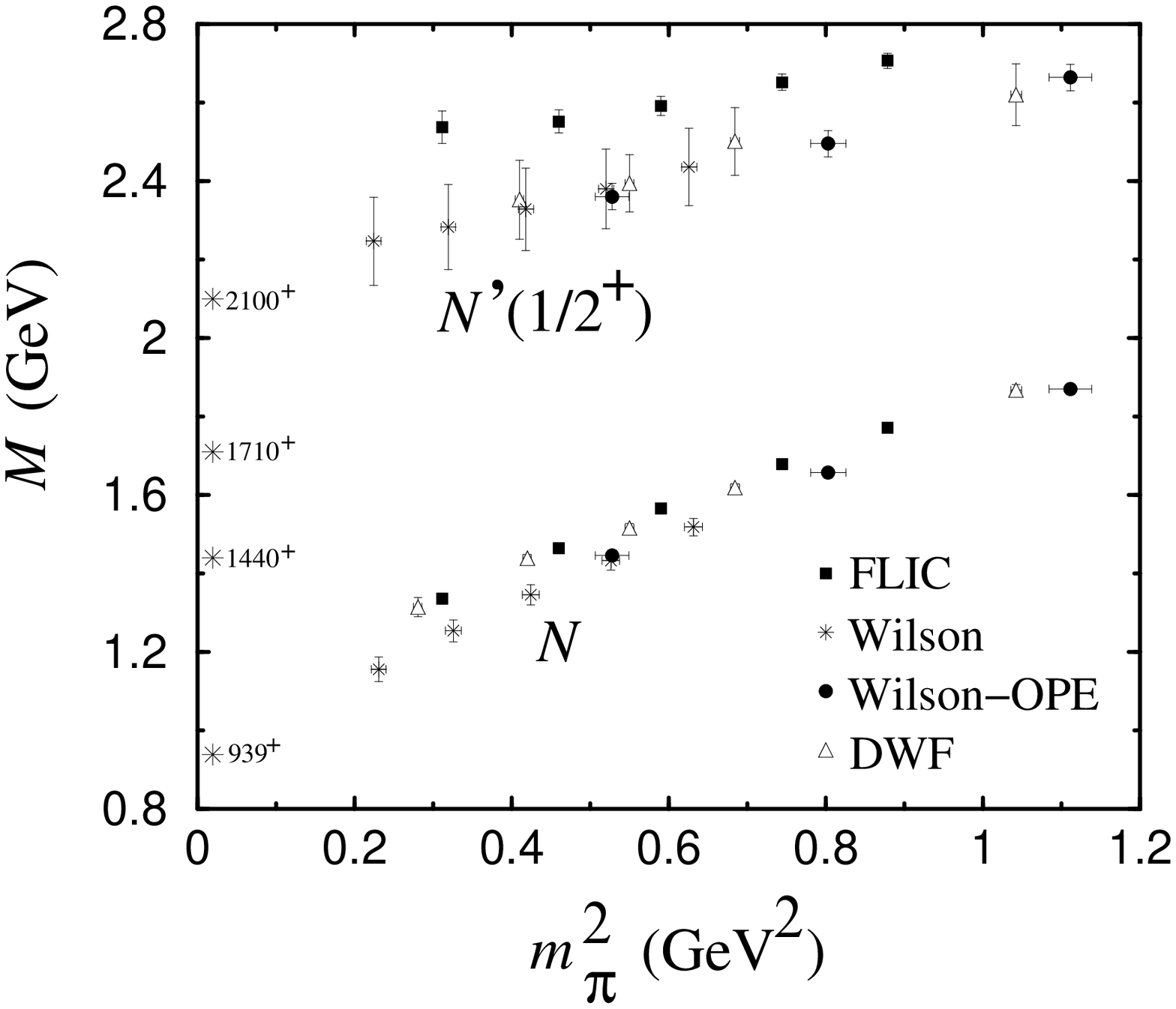,height=10cm}
\vspace*{0.3cm}
\caption{Masses of the nucleon, and the lowest $J^P={1\over 2}^+$
  excitation (``$N'$'').  The FLIC results are compared with the
  earlier DWF \protect\cite{DWF} and Wilson-OPE \protect\cite{LEIN1}
  analyses, as well as with the Wilson results from this analysis.
  The empirical nucleon and low lying $N^*({1\over 2}^+)$ masses are
  indicated by asterisks.}  
\vspace*{0.5cm}
\end{center}
\end{figure*}

\begin{table}[tb!] 
\begin{center}
\caption{Interpolating field coefficients for the two positive parity $N^{1/2^+}$
        states.  The time slice $T$ at which the correlation matrix
        analysis is performed is indicated in the last column.
        \label{CCn}}
\begin{ruledtabular}
  \begin{tabular}{cccccc}
$\kappa$ & $u_1^a$ & $u_2^a$ & $u_1^b$ & $u_2^b$ & $T$ \\ \hline
0.1260 & 0.999(1) & 0.001(1) & 0.154(28) & 0.846(28) & 7 \\ 
0.1266 & 0.997(2) & 0.003(2) & 0.112(59) & 0.888(59) & 8 \\
0.1273 & 0.996(2) & 0.004(2) & 0.083(74) & 0.917(74) & 8  \\
0.1279 & 0.993(3) & 0.007(3) & 0.049(99) & 0.951(99) & 8  \\
0.1286 & 0.989(3) & 0.011(3) & 0.066(81) & 0.934(81) & 7  \\
\end{tabular}
\end{ruledtabular}
\end{center}

\begin{center}
\caption{Interpolating field coefficients for the two negative parity
        $N^{1/2^-}$ states. 
        The eigenvalues of the correlation matrix
        analysis indicate that excited states spoil the eigenstate
        isolation for $\kappa$ values 0.1273 and 0.1279.
        \label{CCn-}}
\begin{ruledtabular}
  \begin{tabular}{cccccc}
$\kappa$ & $u_1^{a*}$ & $u_2^{a*}$ & $u_1^{b*}$ & $u_2^{b*}$ & $T$ \\ \hline
0.1260 & 0.45(7) & 0.55(7) & 0.16(15) & --0.84(15) & 8 \\ 
0.1266 & 0.50(7) & 0.50(7) & 0.08(14) & --0.92(14) & 8 \\
0.1273 & 0.42(6) & 0.58(6) & 0.26(15) & --0.74(15) & 7  \\
0.1279 & 0.53(10) & 0.47(10) & 0.09(15) & --0.91(15) & 7  \\
0.1286 & 0.77(15) & 0.23(15) & 0.20(5) & 0.80(5) & 8  \\
\end{tabular}
\end{ruledtabular}
\end{center}
\end{table}

For comparison, we also show results from earlier simulations with
domain wall fermions (DWF) \cite{DWF} (open triangles), and a
nonperturbatively (NP) improved clover action at $\beta=6.2$
\cite{RICHARDS}.
The scatter of the different NP improved results is due to different
source smearing and volume effects: the open squares are obtained by
using fuzzed sources and local sinks, the open circles use Jacobi
smearing at both the source and sink, while the open diamonds, which
extend to smaller quark masses, are obtained from a larger lattice
($32^3 \times 64$) using Jacobi smearing.
The empirical masses of the nucleon and the three lowest ${1\over 2}^-$
excitations are indicated by the asterisks along the ordinate.

There is excellent agreement between the different improved actions for
the nucleon mass, in particular between the FLIC, DWF \cite{DWF} and
NP improved clover \cite{RICHARDS} results.
On the other hand, the Wilson results lie systematically low in
comparison to these due to the large ${\cal O} (a)$ errors in this
action \cite{FATJAMES}.
A similar pattern is repeated for the $N^*({1\over 2}^-)$ masses.
Namely, the FLIC, DWF and NP improved clover masses are in good
agreement with each other, while the Wilson results again lie
systematically lower.
A mass splitting of around 400~MeV is clearly visible between the $N$
and $N^*$ for all actions, including the Wilson action, despite its poor
chiral properties.
Furthermore, the trend of the $N^*({1\over 2}^-)$ data with decreasing
$m_\pi$ is consistent with the mass of the lowest lying physical
negative parity $N^*$ states.

Figure~6 shows the mass of the $J^P = {1\over 2}^+$ states
(the excited state is denoted by ``$N'(1/2^+)$'').
As is long known, the positive parity $\chi_2$ interpolating field does
not have good overlap with the nucleon ground state \cite{LEIN1} and
the correlation matrix results confirm this result, as discussed below.
It has been speculated that $\chi_2$ may have overlap with the lowest
${1\over 2}^+$ excited state, the $N^*(1440)$ Roper resonance \cite{DWF}.
In addition to the FLIC and Wilson results from the present analysis,
we also show in Fig.~6 the DWF results \cite{DWF}, and results from an
earlier analysis with Wilson fermions together with the operator product
expansion \cite{LEIN1}.
The physical values of the lowest three ${1\over 2}^+$ excitations of
the nucleon are indicated by the asterisks.

The most striking feature of the data is the relatively large excitation
energy of the $N'({1\over 2}^+)$, some 1~GeV above the nucleon.
There is little evidence, therefore, that this state is the $N^*(1440)$
Roper resonance.
While it is possible that the Roper resonance may have a strong nonlinear
dependence on the quark mass at $m_\pi^2 \alt 0.2$~GeV$^2$, arising from,
for example, pion loop corrections, it is unlikely that this behaviour
would be so dramatically different from that of the $N^*(1535)$ so as
to reverse the level ordering obtained from the lattice.
A more likely explanation is that the $\chi_2$ interpolating field does
not have good overlap with either the nucleon or the $N^*(1440)$, but
rather (a combination of) excited ${1\over 2}^+$ state(s).

Recall that in a constituent quark model in a harmonic oscillator basis,
the mass of the lowest mass state with the Roper quantum numbers is
higher than the lowest $P$-wave excitation. It seems that neither the
lattice data (at large quark masses and with our interpolating fields)
nor the constituent quark model have good overlap with the Roper
resonance.
Better overlap with the Roper is likely to require more exotic
interpolating fields.

In Fig.~7 we show the ratio of the masses of the low-lying
$N^*({1\over 2}^-)$ and the nucleon.
Once again, there is good agreement between the FLIC and DWF actions.
However, the results for the Wilson action lie above the others,
as do those for the anisotropic D$_{234}$ action \cite{LEE}.
The D$_{234}$ action has been mean-field improved, and uses an
anisotropic lattice which is relatively coarse in the spatial direction
($a \approx 0.24$~fm).
This is perhaps an indication of the need for nonperturbative or FLIC
improvement.

\begin{figure*}[t] 
\begin{center}
\epsfig{figure=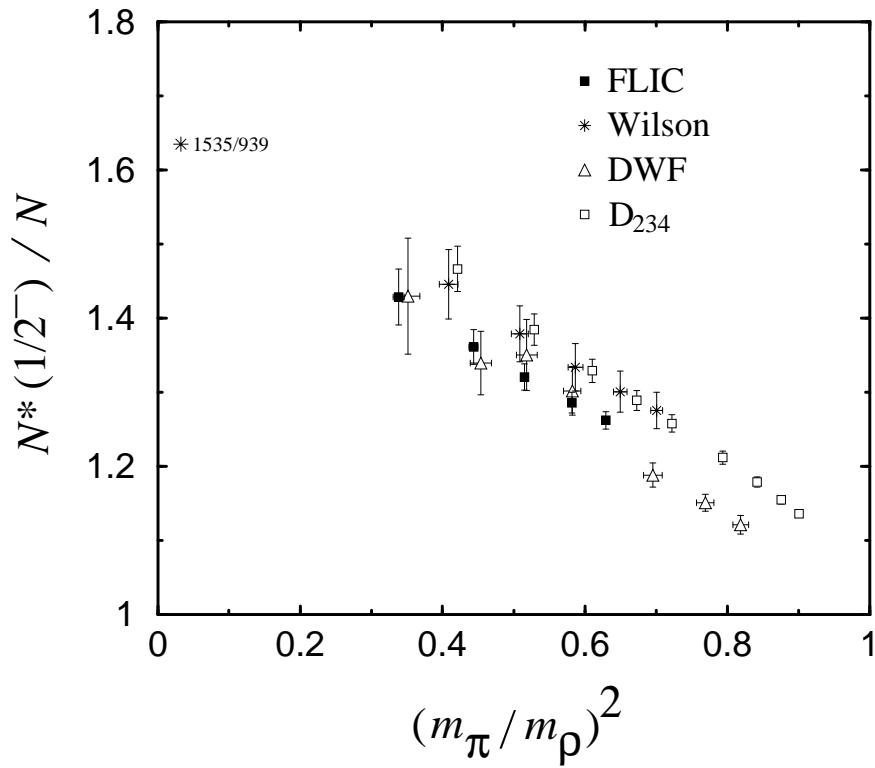,height=10cm}
\vspace*{0.5cm}
\caption{Ratio of the lowest $N^*({1\over 2}^-)$ and nucleon
        masses.  The FLIC and
        Wilson results are from the present analysis, with results from
        the D$_{234}$ \protect\cite{LEE} and DWF \protect\cite{DWF}
        actions shown for comparison.  The empirical $N^*(1535)/N$ mass
        ratio is denoted by the asterisk.}
\vspace*{0.5cm}
\end{center}
\end{figure*}

\subsection{Resolving the resonances}

\begin{figure*}[t] 
\begin{center}
\epsfig{figure=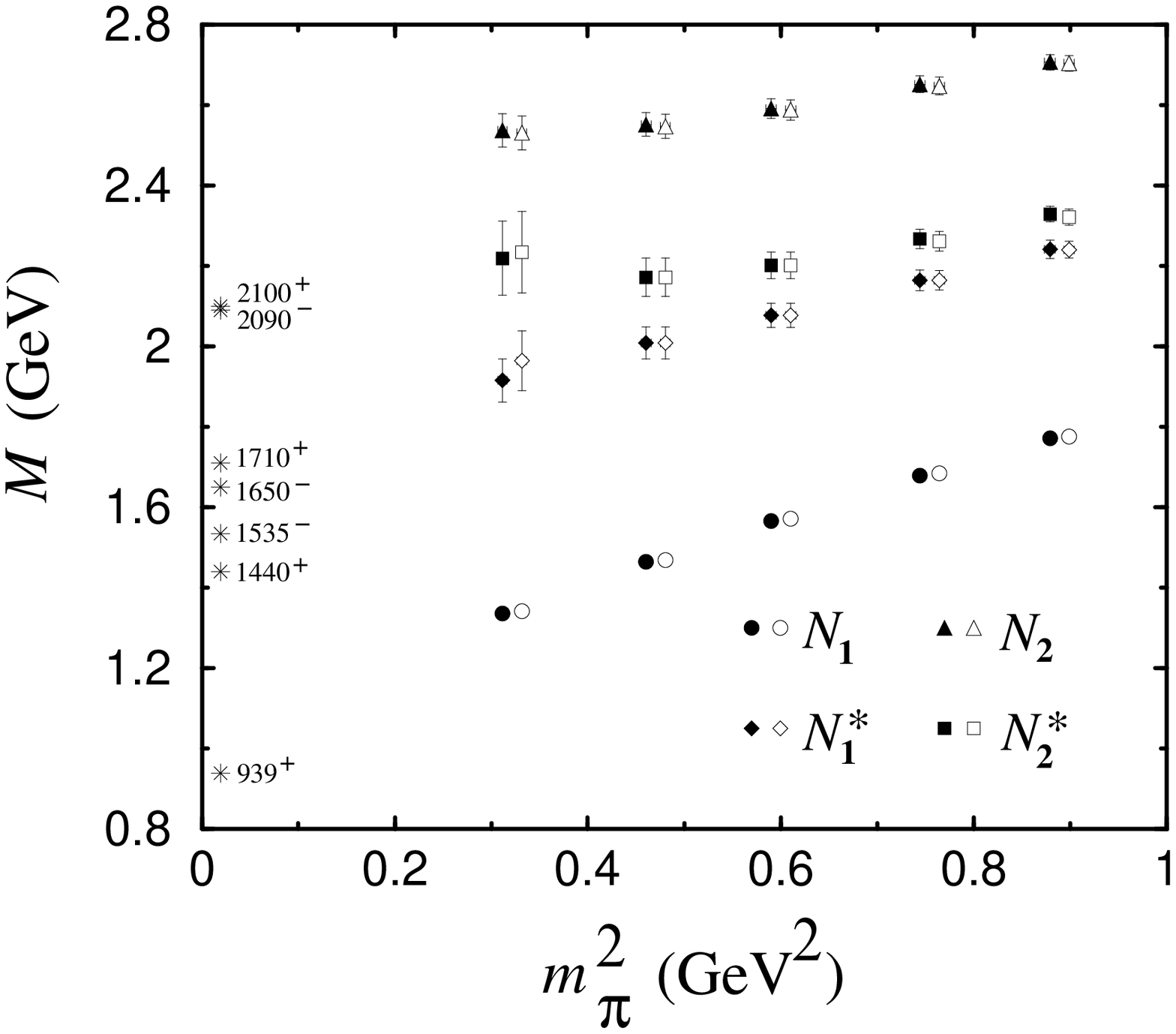,height=10cm}
\vspace*{0.5cm}
\caption{Masses of the $J^P = {1\over 2}^+$ and ${1\over 2}^-$ nucleon
  states, for the FLIC action.  The positive (negative) parity states
  are labeled $N_1$ ($N_1^*$) and $N_2$ ($N_2^*$).  The results from
  the projection of the correlation matrix as discussed in
  Sec.~\ref{6a} are shown by the filled symbols, whereas the results
  from the standard fits to the $\chi_1\overline\chi_1$ and
  $\chi_2\overline\chi_2$ correlation functions are shown by the open
  symbols (offset to the right for clarity).  Empirical masses of the
  low lying ${1 \over 2}^\pm$ states are indicated by the asterisks.
\label{nchipi2}}
\end{center}
\end{figure*}

The mass splitting between the two lightest $N^*({1\over 2}^-)$
states ($N^*(1535)$ and $N^*(1650)$) can be studied by considering the
odd parity content of the $\chi_1$ and $\chi_2$ interpolating fields in
Eqs.~(\ref{chi1p}) and (\ref{chi2p}).
Recall that the ``diquarks'' in $\chi_1$ and $\chi_2$ couple differently
to spin, so that even though the correlation functions built up from the
$\chi_1$ and $\chi_2$ fields will be made up of a mixture of many excited
states, they will have dominant overlap with different states \cite{LEIN1,LL}.
By using the correlation-matrix techniques introduced in the previous section,
we extract two separate mass states from the $\chi_1$ and $\chi_2$
interpolating fields.
The results from the correlation matrix analysis are shown by the
filled symbols in Fig.~8 and are compared to the standard ``naive'' fits
performed directly on the diagonal correlation functions, $\chi_1
\overline{\chi}_1$ and $\chi_2 \overline{\chi}_2$, indicated by the open
symbols.

\begin{table}[tb!] 
\begin{center}
\caption{$\Sigma$ baryon resonance masses.
        \label{sigmadata}}
\begin{ruledtabular}
  \begin{tabular}{ccccc}
        $\kappa$  
      & $m_{\Sigma_1} a$ & $m_{\Sigma^*_1} a$   
      & $m_{\Sigma^*_2} a$ & $m_{\Sigma_2} a$ \\ \hline
    0.1260 & 1.0765(50) & 1.371(15) & 1.432(13) & 1.665(12) \\
    0.1266 & 1.0400(53) & 1.340(16) & 1.404(15) & 1.642(13) \\
    0.1273 & 0.9966(57) & 1.307(17) & 1.371(17) & 1.617(14) \\
    0.1279 & 0.9589(62) & 1.281(20) & 1.349(21) & 1.597(16) \\
    0.1286 & 0.9149(72) & 1.265(29) & 1.332(28) & 1.580(19) \\
\end{tabular}
\end{ruledtabular}
\end{center}
\end{table}

\begin{table}[tb!] 
\begin{center}
\caption{Interpolating field coefficients for the two $\Sigma^{1/2^+}$ states.
        \label{CCsigma}}
\begin{ruledtabular}
  \begin{tabular}{cccccc}
$\kappa$ & $u_1^a$ & $u_2^a$ & $u_1^b$ & $u_2^b$ & $T$ \\ \hline
0.1260 & 0.997(1) & 0.003(1) & 0.127(53) & 0.873(53) & 8 \\ 
0.1266 & 0.997(2) & 0.003(2) & 0.112(59) & 0.888(59) & 8 \\
0.1273 & 0.998(2) & 0.002(2) & 0.133(35) & 0.867(35) & 7  \\
0.1279 & 0.997(2) & 0.003(2) & 0.121(41) & 0.879(41) & 7  \\
0.1286 & 0.996(3) & 0.004(3) & 0.100(52) & 0.900(52) & 7  \\
\end{tabular}
\end{ruledtabular}
\end{center}

\begin{center}
\caption{Interpolating field coefficients for the two $\Sigma^{1/2^-}$ states.
        The eigenvalues of the correlation matrix
        analysis indicate that excited states spoil the eigenstate
        isolation for $\kappa$ values 0.1273 through 0.1286.
        \label{CCsigma-}}
\begin{ruledtabular}
  \begin{tabular}{cccccc}
$\kappa$ & $u_1^{a*}$ & $u_2^{a*}$ & $u_1^{b*}$ & $u_2^{b*}$ & $T$ \\ \hline
0.1260 & 0.47(7) & 0.53(7) & 0.11(13) & --0.89(13) & 8 \\ 
0.1266 & 0.50(7) & 0.50(7) & 0.08(14) & --0.92(14) & 8 \\
0.1273 & 0.38(5) & 0.62(5) & 0.35(14) & --0.65(14) & 7  \\
0.1279 & 0.42(7) & 0.58(7) & 0.30(17) & --0.70(17) & 7  \\
0.1286 & 0.52(13) & 0.48(13) & 0.17(22) & --0.83(22) & 7  \\
\end{tabular}
\end{ruledtabular}
\end{center}
\end{table}

The results indicate that indeed the $N^*({1\over 2}^-)$ largely
corresponding to the $\chi_2$ field (labeled ``$N_2^*$'') lies above
the $N^*({1\over 2}^-)$ which can also be isolated via Euclidean time
evolution with the $\chi_1$ field (``$N_1^*$'') alone.
The masses of the corresponding positive parity states, associated with
the $\chi_1$ and $\chi_2$ fields (labeled ``$N_1$'' and ``$N_2$'',
respectively) are shown for comparison.
For reference, we also list the experimentally measured values of the
low-lying ${1\over 2}^\pm$ states.
It is interesting to note that the mass splitting between the positive
parity $N_1$ and negative parity $N_{1,2}^*$ states (roughly 400--500~MeV)
is similar to that between the $N_{1,2}^*$ and the positive parity $N_2$
state, reminiscent of a constituent quark--harmonic oscillator picture.

\begin{table}[tb!] 
\begin{center}
\caption{$\Xi$ baryon resonance masses.
        \label{xidata}}
\begin{ruledtabular}
  \begin{tabular}{ccccc}
        $\kappa$  
      & $m_{\Xi_1}$  a& $m_{\Xi^*_1 a}$ 
      & $m_{\Xi^*_2} a$ & $m_{\Xi_2} a$ \\ \hline
    0.1260 & 1.0612(52) & 1.358(15) & 1.414(13) & 1.653(12) \\
    0.1266 & 1.0400(53) & 1.340(16) & 1.404(15) & 1.642(13) \\
    0.1273 & 1.0145(54) & 1.320(16) & 1.392(17) & 1.630(14) \\
    0.1279 & 0.9919(56) & 1.302(18) & 1.389(20) & 1.622(15) \\
    0.1286 & 0.9649(60) & 1.281(20) & 1.399(27) & 1.618(24) \\
\end{tabular}
\end{ruledtabular}
\end{center}
\end{table}

The interpolating coefficients for the two positive and negative parity
states (see Eq.~(\ref{lincomIF})), extracted via the procedure outlined
in Section~\ref{recover}, are given in Tables~\ref{CCn} and \ref{CCn-}
for various $\kappa$ values.
The coefficients corresponding to each mass state (labeled ``$a$'' or
``$b$'') are normalised 
\begin{table}[tb!] 
\begin{center}
\caption{Interpolating field coefficients for the two $\Xi^{1/2^+}$ states.
        \label{CCxi}}
\begin{ruledtabular}
  \begin{tabular}{cccccc}
$\kappa$ & $u_1^a$ & $u_2^a$ & $u_1^b$ & $u_2^b$ & $T$ \\ \hline
0.1260 & 0.999(1) & 0.001(1) & 0.146(30) & 0.854(30) & 7 \\ 
0.1266 & 0.997(2) & 0.003(2) & 0.112(59) & 0.888(59) & 8 \\
0.1273 & 0.996(2) & 0.004(2) & 0.105(64) & 0.895(64) & 8  \\
0.1279 & 0.995(2) & 0.005(2) & 0.097(70) & 0.903(70) & 8  \\
0.1286 & 0.993(2) & 0.007(2) & 0.076(83) & 0.924(83) & 8  \\
\end{tabular}
\end{ruledtabular}
\end{center}
%
\begin{center}
\caption{Interpolating field coefficients for the two $\Xi^{1/2^-}$ states.
        The eigenvalues of the correlation matrix
        analysis indicate that excited states spoil the eigenstate
        isolation for $\kappa$ values 0.1273 through 0.1286.
        \label{CCxi-}}
\begin{ruledtabular}
  \begin{tabular}{cccccc}
$\kappa$ & $u_1^{a*}$ & $u_2^{a*}$ & $u_1^{b*}$ & $u_2^{b*}$ & $T$ \\ \hline
0.1260 & 0.48(8) & 0.52(8) & 0.13(16) & --0.87(16) & 8 \\ 
0.1266 & 0.50(7) & 0.50(7) & 0.08(14) & --0.92(14) & 8 \\
0.1273 & 0.38(5) & 0.62(5) & 0.32(13) & --0.68(13) & 7  \\
0.1279 & 0.42(6) & 0.58(6) & 0.22(13) & --0.78(13) & 7  \\
0.1286 & 0.49(7) & 0.51(7) & 0.09(11) & --0.91(11) & 7  \\
\end{tabular}
\end{ruledtabular}
\end{center}
\end{table}
so that the sum of their absolute values is 1,
\begin{subequations}
\begin{eqnarray}
|u_1^a | + |u_2^a | &=& 1\ , \\
|u_1^b | + |u_2^b | &=& 1\ ,
\end{eqnarray}
\end{subequations}
and similarly for the coefficients $u_{1,2}^{a,b\ *}$ for the negative
parity mass states.
This normalisation allows one to readily identify the fraction of each
interpolating field needed to construct a linear combination having
maximum overlap with a particular baryon state.
The last column in Tables~\ref{CCn} and \ref{CCn-} shows the time slice
$T$ where the correlation matrix eigenvalue analysis is performed.

From Table~\ref{CCn} one immediately sees that the coefficient $u_2^a$,
reflecting the fraction of $\chi_2$ required to isolate the ground state nucleon,
is extremely small. This further supports the earlier observation that
the $\chi_2$ interpolating field does not have good overlap with the 
nucleon ground state.
Table~\ref{CCn-} shows the coefficients for isolating the two
lowest-energy negative-parity $N^*$ states using the $\chi_1$ and
$\chi_2$ interpolating fields.
A significant amount of mixing is observed between the two interpolating
fields for the lower energy state, particularly at heavy quark masses.
This result is anticipated by the long Euclidean time evolution
required to achieve an acceptable $\chi^2 / N_{\rm DF}$ for the
$N_1^*$ effective mass illustrated in Fig.~\ref{Neff}.
The higher $N^{1/2^-}$ state, however, is dominated by the $\chi_2$
field, thus explaining the good effective mass plateau observed in
Fig.~\ref{Neff} without the correlation matrix approach.
Note that the most significant contribution to the $N_2^*$ state from
$\chi_1$ is for the third quark mass when the correlation matrix analysis
is performed at an early time slice and is spoiled by
contamination from higher excited states.
The most significant contribution at the preferred time slice, which also 
has the smallest errors, is for the lightest quark mass. 
It is for these reasons that we choose the lightest quark mass in
Fig.~\ref{Neff} to illustrate the effective masses of the projected
nucleon states.


\begin{table}[tb!] 
\begin{center}
\caption{$\Lambda$ baryon resonance masses from the octet,
  $\Lambda^8$, interpolating field.
        \label{lamb8data}}
\begin{ruledtabular}
  \begin{tabular}{ccccc}
        $\kappa$
      & $m_{\Lambda_1} a$ & $m_{\Lambda^*_1} a$ 
      & $m_{\Lambda^*_2} a$ & $m_{\Lambda_2} a$ \\ \hline
    0.1260 & 1.0801(50) & 1.374(15) & 1.427(13) & 1.665(12) \\
    0.1266 & 1.0400(53) & 1.340(16) & 1.404(15) & 1.642(13) \\
    0.1273 & 0.9910(56) & 1.302(17) & 1.380(19) & 1.618(15) \\
    0.1279 & 0.9464(61) & 1.269(21) & 1.373(26) & 1.603(17) \\
    0.1286 & 0.8904(72) & 1.233(28) & 1.410(47) & 1.599(21) \\
\end{tabular}
\end{ruledtabular}
\end{center}
\end{table}

\begin{table}[tb!] 
\begin{center}
\caption{Interpolating field coefficients for the two positive parity
        $\Lambda^8$ states.
        \label{CCl8}}
\begin{ruledtabular}
  \begin{tabular}{cccccc}
$\kappa$ & $u_1^a$ & $u_2^a$ & $u_1^b$ & $u_2^b$ & $T$ \\ \hline
0.1260 & 0.999(1) & 0.001(1) & 0.149(29) & 0.851(29) & 7 \\
0.1266 & 0.997(2) & 0.003(2) & 0.112(59) & 0.888(59) & 8 \\
0.1273 & 0.995(2) & 0.005(2) & 0.095(69) & 0.905(69) & 8  \\
0.1279 & 0.993(2) & 0.007(2) & 0.070(85) & 0.930(85) & 8  \\
0.1286 & 0.990(2) & 0.010(2) & 0.081(63) & 0.919(63) & 7  \\
\end{tabular}
\end{ruledtabular}
\end{center}

\begin{center}
\caption{Interpolating field coefficients for the two negative parity
        $\Lambda^8$ states.
        The eigenvalues of the correlation matrix
        analysis indicate that excited states spoil the eigenstate
        isolation for $\kappa$ values 0.1273 through 0.1286.
        \label{CCl8-}}
\begin{ruledtabular}
  \begin{tabular}{cccccc}
$\kappa$ & $u_1^{a*}$ & $u_2^{a*}$ & $u_1^{b*}$ & $u_2^{b*}$ & $T$ \\ \hline
0.1260 & 0.46(8) & 0.54(8) & 0.16(16) & --0.84(16) & 8 \\ 
0.1266 & 0.50(7) & 0.50(7) & 0.08(14) & --0.92(14) & 8 \\
0.1273 & 0.40(6) & 0.60(6) & 0.27(14) & --0.73(13) & 7  \\
0.1279 & 0.49(8) & 0.51(8) & 0.12(13) & --0.88(13) & 7  \\
0.1286 & 0.47(8) & 0.53(8) & 0.19(13) & --0.81(13) & 6  \\
\end{tabular}
\end{ruledtabular}
\end{center}
\end{table}

\begin{table}[tb!] 
\begin{center}
\caption{$\Lambda$ baryon resonance masses from the ``common'',
  $\Lambda^c$, interpolating field.
        \label{lambcdata}}
\begin{ruledtabular}
  \begin{tabular}{ccccc}
        $\kappa$  
      & $m_{\Lambda_1} a$ & $m_{\Lambda^*_1} a$ 
      & $m_{\Lambda^*_2} a$ & $m_{\Lambda_2} a$ \\ \hline
    0.1260 & 1.0815(50) & 1.334(13) & 1.408(12) & 1.662(11) \\
    0.1266 & 1.0413(52) & 1.301(14) & 1.382(13) & 1.638(12) \\
    0.1273 & 0.9920(56) & 1.262(16) & 1.356(16) & 1.611(12) \\
    0.1279 & 0.9473(61) & 1.226(18) & 1.342(21) & 1.590(13) \\
    0.1286 & 0.8912(73) & 1.181(21) & 1.357(33) & 1.570(15) \\
\end{tabular}
\end{ruledtabular}
\end{center}
\end{table}

\begin{table}[tb!] 
\begin{center}
\caption{Interpolating field coefficients for the two positive parity $\Lambda^c$
        states.
        \label{CClc}}
\begin{ruledtabular}
  \begin{tabular}{cccccc}
$\kappa$ & $u_1^a$ & $u_2^a$ & $u_1^b$ & $u_2^b$ & $T$ \\ \hline
0.1260 & 1.000(2) & 0.000(2) & 0.282(51) & --0.718(51) & 9 \\ 
0.1266 & 0.997(2) & 0.003(2) & 0.291(55) & --0.709(55) & 9 \\
0.1273 & 0.994(2) & 0.006(2) & 0.278(26) & --0.722(26) & 8  \\
0.1279 & 0.990(2) & 0.010(2) & 0.279(18) & --0.721(18) & 7  \\
0.1286 & 0.983(3) & 0.017(3) & 0.278(13) & --0.722(13) & 6  \\
\end{tabular}
\end{ruledtabular}
\end{center}

\begin{center}
\caption{Interpolating field coefficients for the two negative parity $\Lambda^c$
        states.
        The correlation matrix analysis is successful for all $\kappa$
        values.
        \label{CClc-}}
\begin{ruledtabular}
  \begin{tabular}{cccccc}
$\kappa$ & $u_1^{a*}$ & $u_2^{a*}$ & $u_1^{b*}$ & $u_2^{b*}$ & $T$ \\ \hline
0.1260 & 0.54(2) & --0.46(2) & 0.23(3) & 0.77(3) & 8 \\ 
0.1266 & 0.53(2) & --0.47(2) & 0.27(3) & 0.73(3) & 8 \\
0.1273 & 0.52(1) & --0.48(1) & 0.33(3) & 0.67(3) & 8  \\
0.1279 & 0.51(1) & --0.49(1) & 0.39(3) & 0.61(3) & 8  \\
0.1286 & 0.49(1) & --0.51(1) & 0.47(4) & 0.53(4) & 8  \\
\end{tabular}
\end{ruledtabular}
\end{center}
\end{table}

Turning to the strange sector, in Fig.~9 we show the masses of the
positive and negative parity $\Sigma$ baryons calculated from the FLIC
action compared with the physical masses of the known positive and
negative parity states.
The data for the masses of these states are listed in
Table~\ref{sigmadata}, and the interpolator coefficients for the two
positive and negative parity states are given in Tables~\ref{CCsigma}
and \ref{CCsigma-}, respectively.
The pattern of mass splittings is similar to that found in Fig.~8 for
the nucleon.
Namely, the ${1\over 2}^+$ state associated with the $\chi_1$ field
appears consistent with the empirical $\Sigma(1193)$ ground state,
while the ${1\over 2}^+$ state associated with the $\chi_2$ field lies
significantly above the observed first (Roper-like) ${1\over 2}^+$
excitation, $\Sigma^*(1660)$.
There is also evidence for a mass splitting between the two negative parity
states, similar to that in the nonstrange sector.
The behaviour of the interpolator coefficients for the $\Sigma^{1/2^+}$ and
$\Sigma^{1/2^-}$ states is also similar to that for the nucleon in 
Tables~\ref{CCn} and \ref{CCn-}.
Namely, while the positive parity ground state is dominated by the
$\chi_1$ interpolating field, there is considerable mixing between the
$\chi_1$ and $\chi_2$ fields for the lowest negative parity state,
with the higher $\Sigma^{1/2^-}$ state receiving a dominant
contribution from $\chi_2$.

The spectrum of the strangeness --2 positive and negative parity $\Xi$
hyperons is displayed in Fig.~10, with data given in Table~\ref{xidata},
and the interpolator coefficients for the $\Xi^{1/2^+}$ and $\Xi^{1/2^-}$
states in Tables~\ref{CCxi} and \ref{CCxi-}, respectively.
Once again, the pattern of calculated masses repeats that found for
the $\Sigma$ and $N$ masses in Figs.~8 and 9, and for the respective
coupling coefficients.
The empirical masses of the physical $\Xi^*$ baryons are denoted by
asterisks. 
However, for all but the ground state $\Xi(1318)$, the $J^P$
values are not known.
%
%
%

\begin{figure*}[t] 
\begin{center}
\epsfig{figure=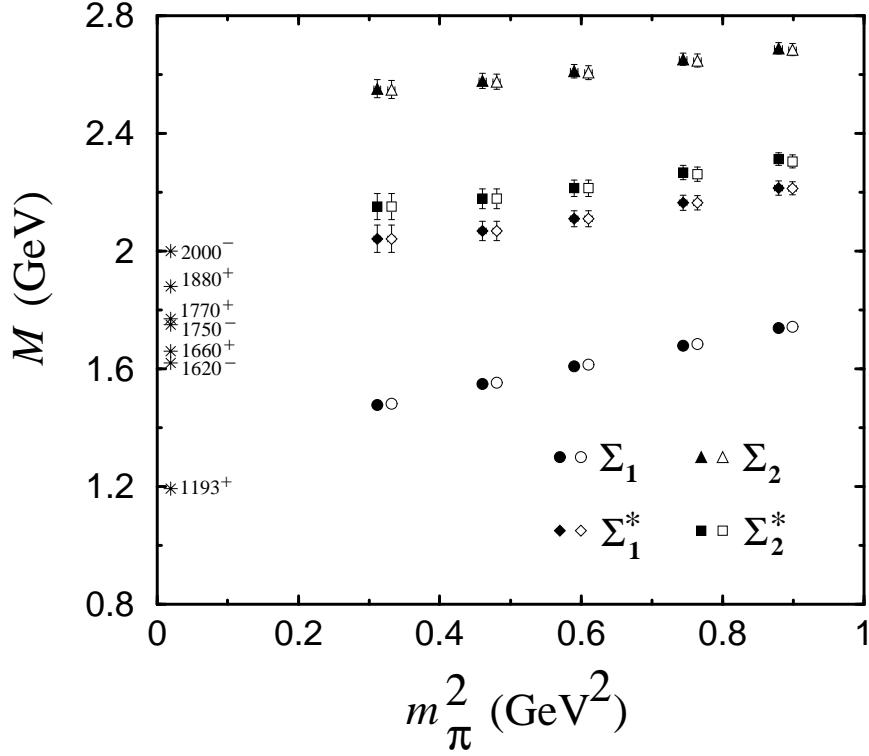,height=10cm}
\vspace*{0.5cm}
\caption{As in Fig.~8 but for the $\Sigma$ baryons.}
\end{center}
\end{figure*}

\begin{figure*}[t] 
\begin{center}
\epsfig{figure=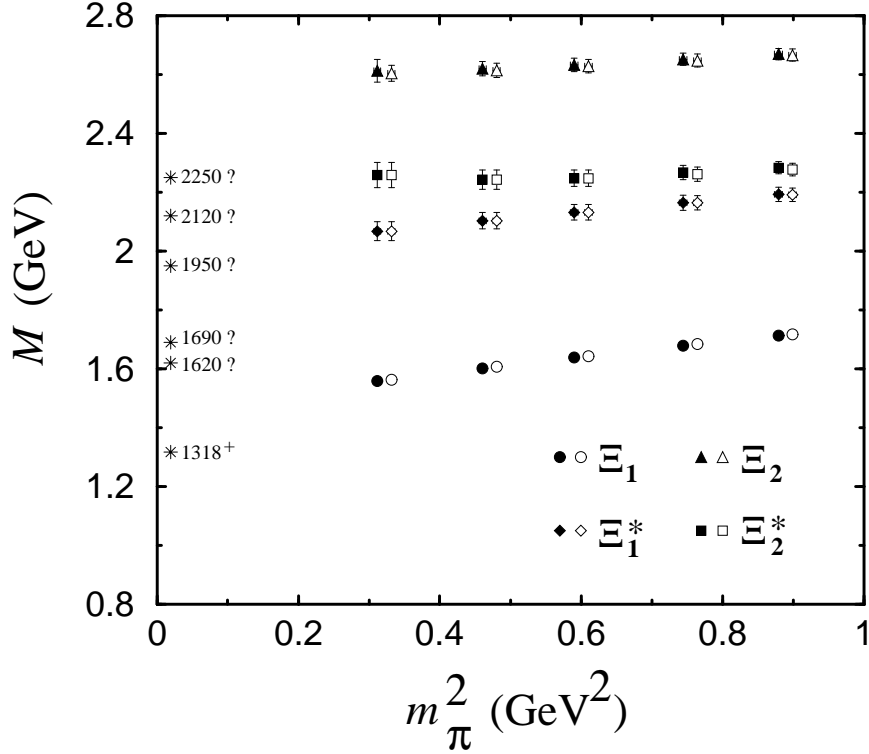,height=10cm}
\vspace*{0.5cm}
\caption{As in Fig.~8 but for the $\Xi$ baryons.
        The $J^P$ values of the excited states marked with
        ``?'' are undetermined.
        \label{Xifig}}
\end{center}
\end{figure*}

\begin{figure*}[t] 
\begin{center}
\epsfig{figure=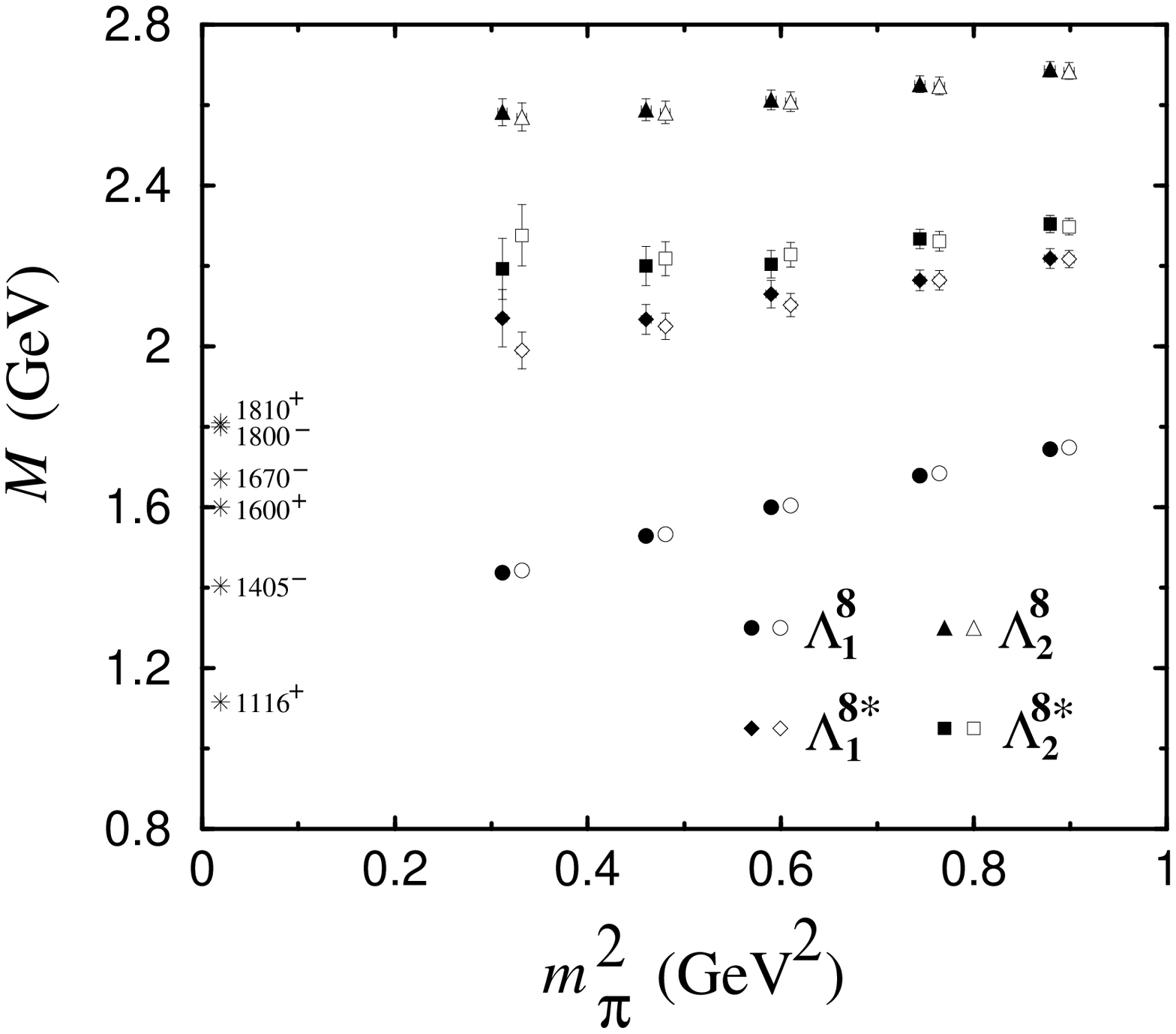,height=10cm}
\vspace*{0.5cm}
\caption{As in Fig.~8 but for the $\Lambda$ states obtained using the
        $\Lambda^8$ interpolating field.
        \label{l8chipi2}}
\end{center}
\end{figure*}

\begin{figure*}[t] 
\begin{center}
\epsfig{figure=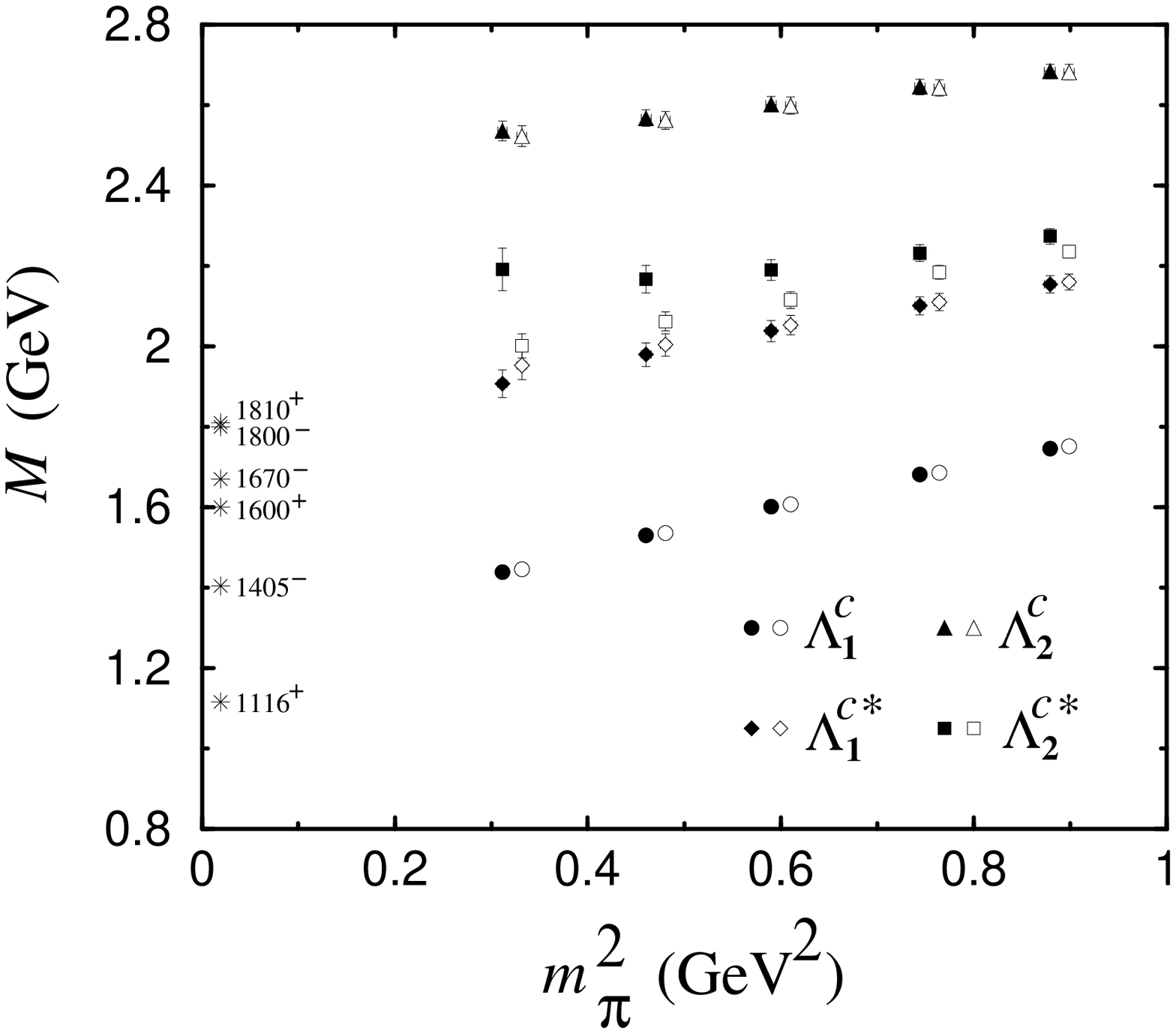,height=10cm}
\vspace*{0.5cm}
\caption{As in Fig.~8 but for the $\Lambda$ states obtained using the
        $\Lambda^c$ interpolating field.
        \label{lcchipi2}}
\end{center}
\end{figure*}

\begin{figure*}[t] 
\begin{center}
\epsfig{figure=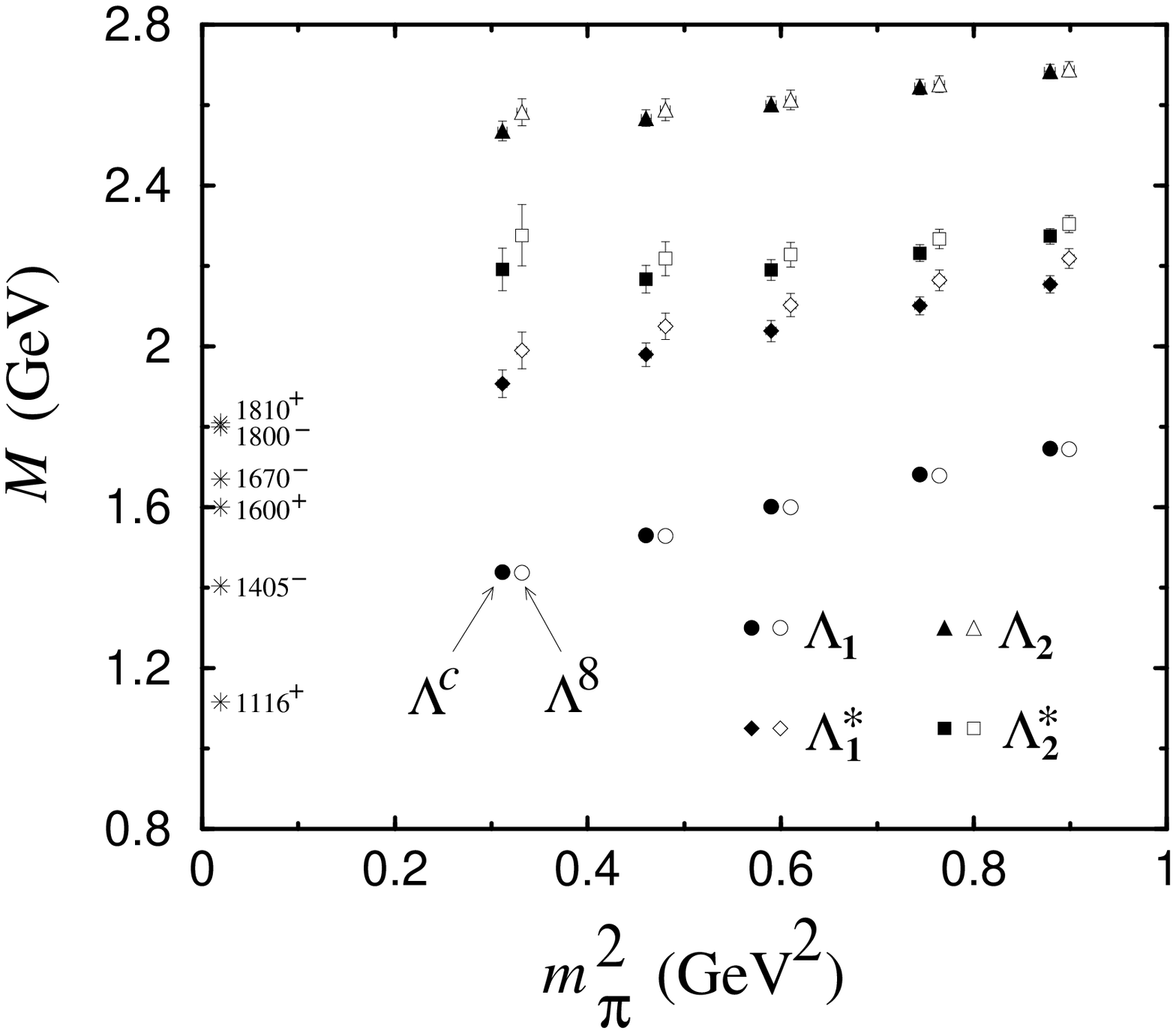,height=10cm}
\vspace*{0.5cm}
\caption{Masses of the positive and negative parity $\Lambda$ states, for
        the octet $\Lambda^8$ (open symbols) and ``common'' $\Lambda^c$
        (filled symbols) interpolating fields with the FLIC action.
        The positive (negative) parity states labeled $\Lambda_1$
        ($\Lambda_1^*$) and $\Lambda_2$ ($\Lambda_2^*$) are the two
        states obtained from the correlation matrix analysis of
        the $\chi_1^\Lambda$ and $\chi_2^\Lambda$ interpolating
        fields.  Empirical masses of the low lying
        ${1\over 2}^\pm$ states are indicated by the asterisks.}
\end{center}
\end{figure*}

Finally, we consider the $\Lambda$ hyperons.  In Figs.~\ref{l8chipi2}
and \ref{lcchipi2} we compare results obtained from the $\Lambda^8$
and $\Lambda^c$ interpolating fields, respectively, using the two
different techniques for extracting masses. The data are given in
Tables~\ref{lamb8data} and \ref{lambcdata}, respectively.
A direct comparison between the positive and negative parity masses for
the $\Lambda^8$ (open symbols) and $\Lambda^c$ (filled symbols) states
extracted from the correlation matrix analysis, is shown in Fig.~13. 
%
%
A similar pattern of mass splittings to that for the
$N^*$ spectrum of Fig.~8 is observed.
In particular, the negative parity $\Lambda_1^*$ state (diamonds) lies
$\sim 400$~MeV above the positive parity $\Lambda_1$ ground state
(circles), for both the $\Lambda^8$ and $\Lambda^c$ fields.
There is also clear evidence of a mass splitting between the 
$\Lambda_1^*$ (diamonds) and $\Lambda_2^*$ (squares).

Using the naive fitting scheme (open symbols in Figs.~\ref{l8chipi2}
and \ref{lcchipi2}),
misses the mass splitting between $\Lambda_1^*$ and $\Lambda_2^*$ for
the ``common'' interpolating field.
Only after performing the correlation matrix analysis is it possible
to resolve two separate mass states, as seen by the filled symbols in
Fig.~\ref{lcchipi2}.
This may be an indication that the physics responsible for the mass
splitting between the negative parity $\Lambda^*(1670)$ and
$\Lambda^*(1800)$ states is suppressed in the $\Lambda^c$
interpolating field. 
This is also evidenced by comparing the interpolator coefficients for the
positive and negative parity $\Lambda^8$ and $\Lambda^c$ states, in
Tables~\ref{CCl8} and \ref{CCl8-}, and \ref{CClc} and \ref{CClc-},
respectively.
While the couplings for the $\Lambda^8$ for both the positive parity
states are similar to those for the nucleon and other hyperons,
there is more prominent mixing for the case of the $\Lambda^c$.
In particular, there is notably stronger mixing for the higher mass
negative parity state in the case of the $\Lambda^c$ compared with
the corresponding $\Lambda^8$ state. The $\chi_2^{\Lambda^8}$
contributes $\sim 80$--90\% of the strength compared to
$\sim 50$--60\% for the $\chi_2^{\Lambda^c}$.
The interpolator coefficients are precisely determined in the
$\Lambda^c$ correlation matrix analysis.
%
As for the other baryons, there is little evidence that the $\Lambda_2$
(triangles) has any significant overlap with the first positive parity
excited state, $\Lambda^*(1600)$ (c.f. the Roper resonance, $N^*(1440)$,
in Fig.~\ref{nchipi2}).

While it seems plausible that nonanalyticities in a chiral extrapolation
\cite{MASSEXTR} of $N_1$ and $N_1^*$ results could eventually lead to
agreement with experiment, the situation for the $\Lambda^* (1405)$ is
not as compelling. 
Whereas a 150~MeV pion-induced self energy is required for the 
$N_1 ,\ N_1^*$ and $\Lambda_1$, 400~MeV is required to approach the 
empirical mass of the $\Lambda^* (1405)$.
This may not be surprising for the octet fields, as the $\Lambda^*(1405)$, 
being an SU(3) flavour singlet, may not couple strongly to an SU(3) octet 
interpolating field. 
Indeed, there is some evidence of this in Fig.~13. 
This large discrepancy of 400~MeV suggests that relevant physics giving rise to a
light $\Lambda^* (1405)$ may be absent from 
simulations in the quenched approximation.
The behaviour of the $\Lambda_{1,2}^*$ states may be modified at small
values of the quark mass through nonlinear effects associated with
Goldstone boson loops including the strong coupling of the
$\Lambda^*(1405)$ to $\Sigma\pi$ and $\bar K N$ channels.
While some of this coupling will survive in the quenched approximation,
generally the couplings are modified and suppressed \cite{YOUNG,SHARPE}.
It is also interesting to note that the $\Lambda_1^*$ and
$\Lambda_2^*$ masses display a similar behaviour to that seen for the
$\Xi_1^*$ and $\Xi_2^*$ states, which are dominated by the heavier
strange quark.
Alternatively, the study of more exotic interpolating fields may indicate
the the $\Lambda^*(1405)$ does not couple strongly to $\chi_1$ or
$\chi_2$.
Investigations at lighter quark masses involving quenched chiral
perturbation theory will assist in resolving these issues.


\section{Conclusion}

We have presented the first results for the excited baryon spectrum from
lattice QCD using an ${\cal O}(a^2)$ improved Luscher-Weise gauge
action \cite{Luscher:1984xn} and an ${\cal O}(a)$-improved
Fat-Link Irrelevant Clover (FLIC) quark action in which only the links of
the irrelevant dimension five operators are smeared \cite{FATJAMES}.
The FLIC action provides a new form of nonperturbative ${\cal O}(a)$
improvement in which ${\cal O}(a)$ errors are eliminated and ${\cal
  O}(a^2)$ errors are very small \cite{QNPproc}.
The simulations have been performed on a $16^3~\times~32$ lattice at
$\beta=4.60$, providing a lattice spacing of $a = 0.122(2)$~fm.
The analysis is based on a set of 400 configurations generated on the 
Orion supercomputer at the University of Adelaide.

Good agreement is obtained between the FLIC and other improved actions,
including the nonperturbatively improved clover \cite{RICHARDS} and domain
wall fermion (DWF) \cite{DWF} actions, for the nucleon and its chiral
partner, with a mass splitting of $\sim 400$~MeV.
Our results for the $N^*({1\over 2}^-)$ improve on those using the
D$_{234}$ \cite{LEE} and Wilson actions.
Despite strong chiral symmetry breaking, the results with the Wilson
action are still able to resolve the splitting between the chiral partners
of the nucleon.
Using the two standard nucleon interpolating fields, we also confirm
earlier observations \cite{LL} of a mass splitting between the two
nearby ${1\over 2}^-$ states.
We find no evidence of overlap with the ${1\over 2}^+$ Roper resonance.

In the strange sector, we have investigated the overlap of various
$\Lambda$ interpolating fields with the low-lying ${1\over 2}^\pm$ states.
Once again a clear mass splitting of $\sim 400$~MeV between the octet
$\Lambda$ and its parity partner is seen, with evidence of a mass
splitting between the two low-lying odd-parity states
We find no evidence of strong overlap with the ${1\over 2}^+$ ``Roper''
excitation, $\Lambda^*(1600)$.
The empirical mass suppression of the $\Lambda^*(1405)$ is not evident in
these quenched QCD simulations, possibly suggesting an important role for
the meson cloud of the $\Lambda^*(1405)$ and/or a need for more exotic
interpolating fields.

We have not attempted to extrapolate the lattice results to the physical
region of light quarks, since the nonanalytic behaviour of $N^*$'s near
the chiral limit is not as well studied as that of the nucleon
\cite{MASSEXTR,YOUNG,Young:2002ib}.
It is vital that future lattice $N^*$ simulations push closer towards the
chiral limit.
On a promising note, our simulations with the 4 sweep FLIC action are able
to reach relatively low quark masses ($m_q \sim 60$--70~MeV) already.
Our discussion of quenching effects is limited to a qualitative level
until the formulation of quenched chiral perturbation theory for
${1\over 2}^-$ baryon resonances is established \cite{CROUCH} or dynamical
fermion simulations are completed.
Experience suggests that dynamical fermion results will be shifted down in
mass relative to quenched results, with increased downward curvature near
the chiral limit \cite{YOUNG}.
It will be fascinating to confront this physics with both numerical
simulation and chiral nonanalytic approaches.

In order to further explore the origin of the Roper resonances or the
$\Lambda^*(1405)$, more exotic interpolating fields involving higher
Fock states, or nonlocal operators should be investigated.
Finally, the present $N^*$ mass analysis will be extended in future to
include $N \to N^*$ transition form factors through the calculation of
three-point correlation functions.

\begin{acknowledgements}

We thank W.~Kamleh, D.G.~Richards, A.W.~Thomas and R.D.~Young for helpful
discussions and communications. 
We would also like to thank Tyson Ritter for his assistance in the
early stages of the correlation matrix investigations.
This work was supported by the Australian Research Council,
and the U.S. Department of Energy contract \mbox{DE-AC05-84ER40150},
under which the Southeastern Universities Research Association (SURA)
operates the Thomas Jefferson National Accelerator Facility
(Jefferson Lab).
The calculations reported here were carried out on the Orion supercomputer
at the Australian National Computing Facility for Lattice Gauge Theory
(NCFLGT) at the University of Adelaide.

\end{acknowledgements}


\end{document}